\documentclass[doublespacing,onecolumn,extra]{gji}
\usepackage{timet}
\usepackage{amsfonts}
\usepackage{mathrsfs}
\usepackage{graphicx}
\usepackage{bm}
\usepackage{caption}
\usepackage{subcaption}
\usepackage{ amssymb }
\usepackage{tikz-cd}
\usepackage{csquotes}

\title[Sea level reciprocity]
      {Reciprocity and sensitivity kernels for  sea level fingerprints}
\author[D. Al-Attar \emph{et al.}]
       {D. Al-Attar$^{1}$,  F. Syvret$^{1}$,  O.  Crawford$^{1}$, J. X. Mitrovica$^{2}$
         and A. J. Lloyd$^{3}$ \\
    $^{1}$ Bullard Laboratories, University of Cambridge, Madingley Road, Cambridge CB3 OEZ, UK.
    Email: da380@cam.ac.uk \\
    $^{2}$Department of Earth and Planetary Sciences, Harvard University, 20 Oxford Street, Cambridge, MA 02138, USA.
    \\
    $^{3}$Lamont Doherty Earth Observatory, Columbia University, Palisades, NY 10964, USA.
  }
\date{Received ?; in original form ?}
\pagerange{\pageref{firstpage}--\pageref{lastpage}}
\volume{142}
\pubyear{2020}
\def\LaTeX{L\kern-.36em\raise.3ex\hbox{{\small A}}\kern-.15em
    T\kern-.1667em\lower.7ex\hbox{E}\kern-.125emX}

\newcommand{\unvec}{\hat{\mathbf{n}}}

\newcommand{\dd}{\,\mathrm{d}}

\newcommand{\bom}{\bm{\omega}}
\newcommand{\Bom}{\bm{\Omega}} 
\newcommand{\vertiii}[1]{{\left\vert\kern-0.25ex\left\vert\kern-0.25ex\left\vert #1 
    \right\vert\kern-0.25ex\right\vert\kern-0.25ex\right\vert}}

\eqsecnum

\begin{document}

\label{firstpage}

\maketitle

\begin{summary}

  Reciprocity theorems are established for the elastic sea level fingerprint problem
  including rotational feedbacks. In  their simplest form, these results
  show that the sea level change at a  location $\mathbf{x}$ due to 
  melting a unit point mass of ice  at $\mathbf{x}'$ is equal to the sea
  level change
  at $\mathbf{x}'$ due to melting a unit point mass of ice  at $\mathbf{x}$.
  This identity holds irrespective of  the shoreline
  geometry or of lateral variations in elastic Earth structure. 
  Using the reciprocity theorems, sensitivity kernels
  for sea level and related observables with respect to the ice load
  can be readily derived. It is notable that calculation of the sensitivity
  kernels is possible using standard fingerprint codes, though
  for some types of observable a slight generalisation to the fingerprint
  problem must be considered. These results are of
  use within coastal hazard assessment and  have a range of potential applications within studies of modern-day sea level change.
\end{summary}


\section{Introduction}

\cite{farrellclark} provided the first gravitationally self-consistent theory of
post-glacial sea level change on a non-rotating, 1-D (depth varying) viscoelastic
Earth for the case of time-invariant shoreline geometry. Their static “sea level
equation” has played a central role in the modern development of the field of glacial
isostatic adjustment (GIA) and, in the intervening half century, their theoretical treatment
has been extended to incorporate 3-D viscoelastic Earth structure and to include both rotational
effects on sea level and shoreline migration due either to local sea level fluctuations or
changes in the perimeter of grounded, marine-based ice \citep[][]{mitrovicamilne,kendall}.

The predictions appearing in \cite{farrellclark} included the special case of elastic Earth
models (e.g., their Figs. 3,4), appropriate for considering the sea level response to ice mass
flux with a time scale shorter than the viscous relaxation time. The calculations highlighted
the counter-intuitive result that sea level falls in the vicinity of a rapidly melting ice sheet
due to the combined effects of elastic crustal uplift and the loss of gravitational attraction
towards the diminishing ice cover. The maximum sea level fall adjacent to the ice sheet is an order
of magnitude (or more) greater than the global mean sea level (GMSL) rise associated with
the melt event, while at large distances from the ice sheet the predicted sea level rise
is up to $\sim 30$\% greater than GMSL. 

\cite{clark1977future} and later \cite{clark1987sea} highlighted the relevance of the elastic case
for predicting geographically variable sea level change arising from ice sheet mass flux driven by
modern global warming. The latter was cited by \cite{mercer1978west} in his canonical study of the
vulnerability of the West Antarctic Ice Sheet to climate change and the implications of its potential
collapse for the Earth system. The connection was further reinforced by \cite{conrad1997spatial} who
quantified the potential bias in estimates of GMSL based on the existing distribution of tide gauge
records. \cite{mitrovica2001recent} were the first to demonstrate that geographic variability in a
subset of tide-gauge determined sea level rates could be reconciled by specific combinations of melt
sources. Their predictions included rotational effects and they emphasised the unique geometry, or
fingerprints \citep{plag2001inversion}, of sea level change associated with each ice sheet or glacier.
The notion of “fingerprinting” the individual sources of melt-water has subsequently become a central
theme in the analysis of modern sea level observations \citep[e.g.][]{tamisiea2001global,milne2009identifying,
  plag2006recent,bamber2010sea,mitrovica2011robustness,brunnabend2015regional,spada2016spectral}. Moreover, the
inclusion of fingerprint physics in such analysis has led to a significant revision in estimates of 20th
century GMSL \citep[][]{hay2015probabilistic,dangendorf2017reassessment}.

In all the above analyses, sea level fingerprints were predicted for specific, assumed geometries of ice
mass flux and this has hindered the adoption of such models in assessments of coastal hazards posed by
a warming world. Three recent studies have inverted these analyses by developing methods for producing a
map that shows the variation in sensitivity of sea level at a specific site to ice mass changes anywhere
within the cryosphere \citep{larour2017should,mitrovica2018quantifying,crawford2018}. These maps represent
sensitivity kernels that can be integrated with arbitrary ice mass flux geometries to compute the sea level
changes at any site of interest. The \cite{larour2017should} approach was based on an automatic differentiation
tool applied to sea level software that solved the forward problem (i.e., computation of the sea level
fingerprint associated with a specific ice mass flux), while Mitrovica et al. (2018) solved a large number
of forward problems in which a grid-by-grid perturbation in ice thickness was prescribed
and applied a finite-difference approximation.
Lastly, \cite{crawford2018} applied the adjoint method to the equations governing
the full sea level problem within a viscoelastic earth model, obtaining results
for elastic fingerprints as a special case. An advantage of this latter method is that it
yields exact sensitivity kernels at a cost equivalent to just two forward calculations (and just one
in the case of the fingerprint problem).
A limitation of that study, along with earlier work on adjoint methods for GIA \citep{alattartromp,martinec2015forward},
was the neglect of rotational feedbacks. Moreover, the  results of  \cite{crawford2018} that are  relevant specifically
to fingerprints are embedded within the more complex viscoelastic theory which makes them harder to understand and apply.

In this article we revisit the derivation of sensitivity kernels through a different and simpler argument that is specialised
to the elastic fingerprint problem  with rotational feedbacks  included and with a fixed shoreline geometry. The latter assumption
renders the fingerprint problem linear and  is appropriate within applications to modern sea level where amplitudes of change are low.
Our presentation is deliberately independent of
\cite{crawford2018}, and hence a knowledge of that work and the Lagrange multiplier methods on which it is based is not necessary to understand this paper.
We proceed here by first establishing  reciprocity theorems for the fingerprint problem along with suitable generalisations thereof.
In their simplest form,
these results show the following: Suppose that at some location, $\mathbf{x}$,
a unit  point mass of ice is melted and the sea level change is observed at $\mathbf{x}'$. This sea level change
is identical to that that would be observed at $\mathbf{x}$ due to melting a unit point mass of ice at $\mathbf{x}'$. The derivation makes no assumptions about the shoreline geometry nor internal structure of the earth model,
and hence this is not merely a geometric symmetry of the problem.
With the reciprocity theorems at hand, it is a simple matter to obtain sensitivity kernels for
the whole range of physically relevant observables. In the case of observables depending directly on
sea level change (e.g. tide gauge determined rates), calculation of the sensitivity kernel can be
performed using a standard fingerprint code. When more general observables are considered
(e.g., those associated with satellite altimetry or gravity) 
a  generalised form of the fingerprint problem must be solved, but the necessary changes are slight. While the
focus of this work is on  sea level change driven by the growth or melting of continental ice sheets, it is worth commenting
that the theory developed is applicable to other types of surface loading  such as that associated with hydrology, ocean
dynamics or sedimentation.

\section{Sensitivity kernels in non-rotating earth models}

\label{sec:nonrot}

Within this section we derive reciprocity theorems for the elastic fingerprint problem  in a non-rotating earth
model and show how they can be used to obtain sensitivity kernels for sea level (and other observables) with
respect to ice thickness. The necessary extensions to account for rotational feedbacks are described in
Section \ref{sec:rot}. Splitting the theory in this manner is unnecessary but we think it helpful to incrementally add complicating factors
into the problem.

\subsection{Equations of motion for static loading}

We start by considering the static loading of a  non-rotating elastic earth model.
The volume of the earth model will be written $M$ and its surface $\partial M$.
It is assumed that, prior to application of the load, the earth model is in
a state of hydrostatic equilibrium, with its gravitational potential
written $\Phi$. As discussed in Section 2.3.2 of \cite{alattartromp}, the response
of such a model to a surface load, $\sigma$, is governed by the following equations
of motion expressed in the weak form
\begin{equation}
  \label{eq:eqmnr}
  \mathcal{A}(\mathbf{u},\phi\,|\, \mathbf{u}',\phi')
  + \int_{\partial M} (\mathbf{u}'\cdot \nabla \Phi + \phi')\,\sigma \dd S = 0. 
\end{equation}
Here $\mathbf{u}$ is the displacement vector, $\phi$ the Eulerian gravitational
potential perturbation, while $\mathbf{u}'$ and $\phi'$ are corresponding
test functions. We recall that within the weak formulation, the fields $(\mathbf{u},\phi)$
solve the problem if the above equation holds for arbitrary test functions
(subject to standard regularity conditions). The term
$\mathcal{A}(\mathbf{u},\phi\,|\, \mathbf{u}',\phi')$, which is associated with the elastic and gravitational response of the
earth model, is a bilinear form whose
definition can be found in eq.(2.52) of \cite{alattartromp}. For
our purposes, we need only  the following two properties of this bilinear form:
\begin{enumerate}
\item{ It is symmetric in the sense that
  \begin{equation}
    \mathcal{A}(\mathbf{u},\phi\,|\, \mathbf{u}',\phi') =
    \mathcal{A}( \mathbf{u}',\phi' \,|\, \mathbf{u},\phi).
  \end{equation}
  Note that within the identity the displacement and gravitational
  potential perturbations are paired together; we cannot, for example,
  interchange the two displacements vectors alone.}
\item{ For any test functions $(\mathbf{u}',\phi')$, the following identity holds
  \begin{equation}
    \mathcal{A}(\mathbf{u},\phi\,|\, \mathbf{u}',\phi') = 0
  \end{equation}
  when the displacement and gravitational potential perturbation take the form
  \begin{equation}
    \mathbf{u} = \mathbf{a} + \mathbf{b}\times \mathbf{x}, \quad
    \phi = -(\mathbf{a} + \mathbf{b}\times \mathbf{x}) \cdot \nabla\Phi,
  \end{equation}
  with arbitrary constant vectors $\mathbf{a}$ and $\mathbf{b}$.
  The fields $\mathbf{u}$ and $\phi$ defined here correspond  to a linearised rigid body motion, with
  the vector $\mathbf{a}$ parameterising the translational degrees of freedom and $\mathbf{b}$
those of rotation.}
\end{enumerate}
Using these two properties, we can take 
\begin{equation}
  \mathbf{u}' = \mathbf{a} + \mathbf{b}\times \mathbf{x}, \quad
  \phi' = -(\mathbf{a} + \mathbf{b}\times \mathbf{x}) \cdot \nabla\Phi,
\end{equation}
within eq.(\ref{eq:eqmnr}) to arrive at a trivial equality. From this it follows
via the Fredholm alternative \citep[e.g][Chapter 6]{marsden1994mathematical} that the static loading
problem  admits a solution for any $\sigma$, but that the solution
is only defined up to  a linearised rigid body motion \citep[c.f.][Section 3]{martinec2015forward}.

\subsection{A first reciprocity theorem}

Suppose that $(\mathbf{u},\phi)$ solve the static loading problem
for a load $\sigma$, and that $(\mathbf{u}^{\dagger},\phi^{\dagger})$ solve the static loading problem
for some other load $\sigma^{\dagger}$. From eq.(\ref{eq:eqmnr}), we know that
\begin{equation}
  \mathcal{A}(\mathbf{u},\phi\,|\, \mathbf{u}^{\dagger},\phi^{\dagger})
  + \int_{\partial M} (\mathbf{u}^{\dagger}\cdot \nabla \Phi + \phi^{\dagger})\,\sigma \dd S = 0.
\end{equation}
In exactly the same way for the second problem, we can write
\begin{equation}
  \mathcal{A}(\mathbf{u}^{\dagger},\phi^{\dagger}\,|\, \mathbf{u},\phi)
  + \int_{\partial M} (\mathbf{u}\cdot \nabla \Phi + \phi)\,\sigma^{\dagger} \dd S = 0.
\end{equation}
Subtracting one equality from the other and using the symmetry of $\mathcal{A}$,
we arrive at the non-trivial identity
\begin{equation}
  \label{eq:rec1}
  \int_{\partial M} (\mathbf{u}^{\dagger}\cdot \nabla \Phi + \phi^{\dagger})\,\sigma \dd S =
  \int_{\partial M} (\mathbf{u}\cdot \nabla \Phi + \phi)\,\sigma^{\dagger} \dd S.
\end{equation}
This is not a new result. It was obtained by \cite{trompmitrovica}
within the context of quasi-static loading problems and is, as they
pointed out,  a generalisation of Betti's reciprocity theorem that is well-known in seismology
\citep[e.g.][]{akirichards}. 

\subsection{Reformulating the theorem in terms of sea level}

Within gravitationally self-consistent sea level theory
\citep[e.g.][]{farrellclark,mitrovicamilne,crawford2018},  deformation of the earth model
can be related to sea level change, $\Delta SL$, through the relation
\begin{equation}
  \label{eq:sldef}
  \Delta SL = -\frac{1}{g}(\mathbf{u}\cdot \nabla \Phi + \phi) + \frac{\Phi_{g}}{g},
\end{equation}
where $g$ is the magnitude of gravitational acceleration at the surface. Here we note that
we are working with a hydrostatic theory in which 
sea level is defined as  difference been the equipotential that defines the sea surface and the solid surface
\citep[e.g.][]{mitrovicamilne,tamisiea2011ongoing}. The first term
on the right hand side of eq.(\ref{eq:sldef}) is associated with vertical motion of the solid surface, the
second with changes in the shape of equipotential surfaces, while the spatially constant final 
term, $\frac{\Phi_{g}}{g}$, is linked to variations in ocean mass. As a comment on notations, within
this paper, we consider sea level change, $\Delta SL$,
along with related quantities  such
as ice thickness change, or global mean sea level change that are defined later.
Because, however, all equations are linear and have
no explicit time-dependence, we could equally well have chosen to work with with rates of change
of these quantities.
The value of the  constant $\Phi_{g}$ is fixed by
conserving mass between the oceans and direct load,
with this requirement conveniently written as
\begin{equation}
  \int_{\partial M} \sigma \dd S = 0.
\end{equation}
Here  the total  load, $\sigma$, is decomposed into
an ocean load and a direct load, $\zeta$, as
\begin{equation}
  \label{eq:loaddec}
  \sigma = \rho_{w} C \, \Delta SL + \zeta, 
\end{equation}
where $\rho_{w}$ is the density of water and $C$ the ocean function
that equals one where water is present and zero otherwise.
In the case of ice loading, the direct term is given by
\begin{equation}
  \zeta = \rho_{i} (1-C) \Delta I, 
\end{equation}
where $\rho_{i}$ is the density of ice, and $\Delta I$ the change in
ice thickness. The factor, $1-C$, within this expression accounts
for the possibility of floating ice \citep[e.g.][equations 31--34]{crawford2018}.

Consider again  a pair of solutions $(\mathbf{u},\phi)$
and $(\mathbf{u}^{\dagger},\phi^{\dagger})$ of the loading problem associated, respectively, to with loads $\sigma$ and $\sigma^{\dagger}$.
Here, however, we  assume that these loads are  decomposed
as in eq.(\ref{eq:loaddec}) into water  and direct terms that share a common ocean function.
From the above expression for sea level change we can write
\begin{equation}
  \mathbf{u}\cdot \nabla \Phi + \phi = - g \, \Delta SL + \Phi_{g},  \quad
  \mathbf{u}^{\dagger}\cdot \nabla \Phi + \phi^{\dagger} = - g \, \Delta SL^{\dagger} + \Phi_{g}^{\dagger},  
\end{equation}
and hence eq.(\ref{eq:rec1}) becomes
\begin{equation}
  \int_{\partial M} (- g \, \Delta SL^{\dagger} + \Phi^{\dagger}_{g})\,\sigma \dd S =
  \int_{\partial M} ( - g \, \Delta SL + \Phi_{g} )\,\sigma^{\dagger} \dd S.
\end{equation}
The terms involving the constants $\Phi_{g}$ and $\Phi_{g}^{\dagger}$ vanish due to
conservation of mass, and so the identity simplifies to
\begin{equation}
  \int_{\partial M} \Delta SL^{\dagger} \,\sigma \dd S =
  \int_{\partial M}  \Delta SL \,\sigma^{\dagger} \dd S.
\end{equation}
If we now substitute into the equality the decompositions of the loads,
$\sigma$ and $\sigma^{\dagger}$, we find
\begin{equation}
  \int_{\partial M} \Delta SL^{\dagger} (\rho_{w} C \, \Delta SL + \zeta) \, \dd S =
  \int_{\partial M}  \Delta SL \,(\rho_{w} C \, \Delta SL^{\dagger} + \zeta^{\dagger}) \dd S, 
\end{equation}
and cancelling the terms symmetric in $\Delta SL$ and $\Delta SL^{\dagger}$ we arrive at 
\begin{equation}
  \label{eq:rec2}
  \int_{\partial M} \Delta SL^{\dagger} \, \zeta \, \dd S =
  \int_{\partial M}  \Delta SL \, \zeta^{\dagger} \dd S.
\end{equation}
Again, this is a known result, being implied as a special case by the
adjoint theory of \cite{crawford2018} for sea level change
in a viscoelastic earth model. What  is
new  is the explicit statement as a reciprocity theorem
along with the more elementary derivation that
has been facilitated by restricting attention to
the  elastic fingerprint problem. 

\subsection{Symmetry of the Green's function}

Because the fingerprint problem is linear, its solution must take the form
\begin{equation}
  \Delta SL(\mathbf{x}) = \int_{\partial M} G(\mathbf{x},\mathbf{x}') \zeta(\mathbf{x}') \dd S_{\mathbf{x}'}, 
\end{equation}
for an appropriate Green's function, where we have added a subscript to the surface element to make clear which
variable it is defined with respect to. Suppose that, within eq.(\ref{eq:rec2}), we take
\begin{equation}
  \zeta^{\dagger}(\mathbf{x}') = \delta(\mathbf{x}',\mathbf{x}), 
\end{equation}
with $\delta(\mathbf{x}',\mathbf{x}) $ being the delta function
on $\partial M$ based at $\mathbf{x}$ which is defined such that
\begin{equation}
  \int_{\partial M} f(\mathbf{x}') \delta(\mathbf{x}',\mathbf{x}) \dd S_{\mathbf{x}'} =
  f(\mathbf{x}),
\end{equation}
for any suitably regular function, $f$.
By definition of the Green's function, it follows that $\Delta SL^{\dagger}(\mathbf{x}') = G(\mathbf{x}',\mathbf{x})$, and hence
from the reciprocity theorem we arrive at the identity
\begin{equation}
  G(\mathbf{x},\mathbf{x}') =   G(\mathbf{x}',\mathbf{x}).
\end{equation}
Such a symmetry of the Green's function was established for static loading problems  by \cite{trompmitrovica}
in cases where there is only a direct load. Here we see the result  remains
true when a  gravitationally self-consistent water load is included.  It is this
symmetry that  makes precise our earlier statements
of the reciprocity theorem
in terms of the response to  melting unit point masses of ice.

\subsection{Self-adjointness of the sea level equation}

An alternative description of the sea level reciprocity theorem can be given in the language of functional analysis.
Solution of the the sea level equation  implicitly defines a linear mapping, $A$, from the direct load, $\zeta$,
to the resulting sea level change, $\Delta SL$, with both scalar fields being defined on the surface, $\partial M$,
of the earth model. Using this notation  within eq.(\ref{eq:rec2}), we have
\begin{equation}
  \int_{\partial M} (A \zeta^{\dagger}) \,\zeta \dd S =   \int_{\partial M} (A \zeta) \,\zeta^{\dagger} \dd S.
\end{equation}
Regarding the integral, $\int_{\partial M} u v \dd S$, of two real-valued scalar fields, $u$ and $v$, on $\partial M$ as an
inner product, it follows that the operator, $A$, is self-adjoint \citep[e.g.][]{schechter2001principles}.
To make these ideas precise, one should  specify the domain and range of the operator, $A$.
We will not dwell on such technical points nor pursue this operator formalism further within this work, but  note that these
ideas can also be  applied to the generalised  reciprocity theorems discussed below.

\subsection{Sensitivity kernels for sea level change}

Suppose now that $J(\Delta SL)$ is some functional of the sea level change. For example,
within the context of an inverse problem to recover the direct load
we might consider
\begin{equation}
  J(\Delta SL) = \frac{1}{2}\int_{\partial M} C \,(\Delta SL-\Delta SL_{\mathrm{o}})^{2} \dd S, 
\end{equation}
where $\Delta SL_{\mathrm{o}}$ is the observed sea level change and the ocean function is included
to limit contributions to the oceans. Because
$\Delta SL$ is obtained by solving the sea level problem for a given direct load, $\zeta$,
this functional  implicitly depends on $\zeta$. It can,
for a variety of reasons, be interesting to determine the functional derivative
of $J$ with respect to $\zeta$. To do so, we write $\delta \zeta$
for the perturbation in the direct load, and $\delta \Delta SL$ for the
corresponding perturbation to $\Delta SL$. Working to first-order in
perturbed quantities, we have 
\begin{equation}
  \delta J  = \int_{\partial M} C\, (\Delta SL-\Delta SL_{\mathrm{o}})\, \delta \Delta SL \dd S, 
\end{equation}
for the above example, or more generally
\begin{equation}
  \delta J  = \int_{\partial M} \zeta^{\dagger}\, \delta \Delta SL \dd S, 
\end{equation}
for some appropriate $\zeta^{\dagger}$ determined by the choice of $J$.
Because the fingerprint problem is linear, eq.(\ref{eq:rec2})
immediately holds with $(\Delta SL,\zeta)$ replaced by $(\delta \Delta SL,\delta \zeta)$,
and hence the reciprocity theorem then gives 
\begin{equation}
  \label{eq:kernel_zeta}
  \delta J = \int_{\partial M} \Delta SL^{\dagger} \,\delta \zeta \dd S, 
\end{equation}
where $\Delta SL^{\dagger}$ is obtained by solving the fingerprint problem
subject to the direct load $\zeta^{\dagger}$.
This shows  that $\Delta SL^{\dagger}$ is the  sensitivity kernel of $J$
with respect to $\zeta$. Recalling that in the case of
an ice load we have $\zeta = \rho_{i}(1-C) \Delta I$, it follows
that we can  equivalently write
\begin{equation}
  \label{eq:kernel_ice}
  \delta J = \int_{\partial M} \rho_{i}\, (1-C) \,\Delta SL^{\dagger} \,\delta \Delta I \dd S, 
\end{equation}
from which we can identify the sensitivity kernel of $J$ with respect
to the ice thickness
\begin{equation}
  K = \rho_{i}\, (1-C)\, \Delta SL^{\dagger}.
\end{equation}
These results are special cases of  those obtained by \cite{crawford2018}. Following the terminology
of that work, we will call $\Delta SL^{\dagger}$  the adjoint sea level
and $\zeta^{\dagger}$ the adjoint load.
What this new derivation makes clear is that  sensitivity kernels can be  obtained by solving the standard
fingerprint problem subject to an appropriate adjoint load.

\subsection{Dealing with more complex observables}

We have now seen that sensitivity kernels can be readily obtained for  functionals, $J$,
defined in terms of the sea level change, $\Delta SL$.
In practice there are, of course, other quantities that can be observed,
including changes in sea surface height or in the gravitational potential.
To obtain analogous results for such observables, it is necessary to consider the
generalised loading problem
\begin{equation}
  \label{eq:eqmnrg}
  \mathcal{A}(\mathbf{u},\phi\,|\, \mathbf{u}',\phi')
  + \int_{\partial M} (\mathbf{u}'\cdot \nabla \Phi + \phi')\,\sigma \dd S
  + \int_{\partial M} (\mathbf{u}'\cdot \mathbf{t} + \zeta_{\phi} \phi') \dd S = 0,
\end{equation}
where we have introduced additional forcing terms  $\mathbf{t}$ and $\zeta_{\phi}$. These
terms need not have a direct physical interpretation, but we will see that
the term $\mathbf{t}$ allows
us to consider functionals depending directly on the surface displacement, while $\zeta_{\phi}$
is needed for those involving the gravitational potential perturbation.
Recalling the second property of the bilinear form
mentioned above, it follows that a necessary and sufficient condition for solutions to exist is
\begin{equation}
  \int_{\partial M} (\mathbf{t} - \zeta_{\phi} \nabla \Phi)\cdot  (\mathbf{a} + \mathbf{b}\times \mathbf{x})
  \dd S = 0,
\end{equation}
with $\mathbf{a}$ and $\mathbf{b}$ arbitrary constant vectors. The practical significance of this
condition will be explained below. 

A reciprocity theorem for the generalised loading problem
can be  obtained  as before by introducing
pairs of solutions  $(\mathbf{u},\phi)$
and $(\mathbf{u}^{\dagger},\phi^{\dagger})$ corresponding, respectively, to the generalised loads
$(\sigma,\mathbf{t},\zeta_{\phi})$ and $(\sigma^{\dagger},\mathbf{t}^{\dagger},\zeta^{\dagger}_{\phi})$.
The result is the following identity
\begin{equation}
  \label{eq:rec3_tmp}
  \int_{\partial M} (\mathbf{u}^{\dagger}\cdot \nabla \Phi + \phi^{\dagger})\,\sigma \dd S
  + \int_{\partial M} (\mathbf{t} \cdot \mathbf{u}^{\dagger} + \zeta_{\phi} \phi^{\dagger}) \dd S
  = \int_{\partial M} (\mathbf{u}\cdot \nabla \Phi + \phi)\,\sigma^{\dagger} \dd S
  + \int_{\partial M} (\mathbf{t}^{\dagger}\cdot \mathbf{u} + \zeta^{\dagger}_{\phi} \phi) \dd S.
\end{equation}
By decomposing the loads $\sigma$ and $\sigma^{\dagger}$ into
direct terms and matching water loads, the above identity  becomes
\begin{equation}
  \label{eq:rec3}
  \int_{\partial M} \Delta SL^{\dagger} \,\zeta \dd S
  -\frac{1}{g} \int_{\partial M} (\mathbf{t}\cdot \mathbf{u}^{\dagger}  + \zeta_{\phi} \phi^{\dagger}) \dd S
  = \int_{\partial M} \Delta SL \, \zeta^{\dagger} \dd S
  -\frac{1}{g} \int_{\partial M} (\mathbf{t}^{\dagger}\cdot \mathbf{u} + \zeta^{\dagger}_{\phi} \phi) \dd S, 
\end{equation}
which is a reciprocity theorem for the generalised fingerprint problem. 

Suppose  that we define some functional $J(\Delta SL,\mathbf{u},\phi)$
in terms of the solution to the fingerprint problem and wish to determine
its functional derivative with respect to the direct load, $\zeta$. 
To first-order, the perturbation in $J$ can be written
\begin{equation}
  \label{eq:DJ}
  \delta J = \int_{\partial M} \delta \Delta SL\,  \zeta^{\dagger}   \dd S
  + \int_{\partial M} (\mathbf{t}^{\dagger}\cdot \delta \mathbf{u} +
  \zeta^{\dagger}_{\phi} \delta \phi) \dd S, 
\end{equation}
for suitable choices of $(\zeta^{\dagger},\mathbf{t}^{\dagger},\zeta_{\phi}^{\dagger})$.
Eq.(\ref{eq:rec3}) then
leads immediately to the desired relation
\begin{equation}
  \delta J = \int_{\partial M} \Delta SL^{\dagger} \,\delta \zeta \dd S, 
\end{equation}
with  the adjoint sea level, $\Delta SL^{\dagger}$ obtained through
the solution of the generalised fingerprint problem subject to the
generalised loads $(\sigma^{\dagger},\mathbf{t}^{\dagger},\zeta^{\dagger}_{\phi})$
defined within eq.(\ref{eq:DJ}).   Here we must ask whether the problem for $\Delta SL^{\dagger}$
has a well-defined solution, this requiring
\begin{equation}
  \int_{\partial M} (\mathbf{t}^{\dagger} - \zeta_{\phi}^{\dagger} \nabla \Phi)\cdot  (\mathbf{a} + \mathbf{b}\times \mathbf{x})
  \dd S = 0,
\end{equation}
for any constant vectors $\mathbf{a}$ and $\mathbf{b}$. Clearly, however, this condition is equivalent to the
requirement that  the value of $J$
is invariant under linearised rigid body motions.
Equally, the solution to the generalised fingerprint problem is 
only defined up to a linearised rigid body motion, but
such a  term makes no contribution to the resulting sea level change, $\Delta SL^{\dagger}$.
The end result is that the sensitivity kernels are well-defined
so long as we work  with observables that
are physically meaningful.

\section{Sensitivity kernels in rotating earth models}

\label{sec:rot}

Having established the basic methods in a non-rotating earth model, we show in this section
how rotational feedbacks can be incorporated into the problem. Doing this requires a  brief
review of  rotational dynamics for a deforming planet, but once this is done  we can
proceed very quickly to the final results.

\subsection{Review of some rotational dynamics}

 Consider an earth model
rotating steadily with angular velocity, $\Bom$.
The steady-state Euler equation takes the form
\begin{equation}
  \Bom \times (\mathbf{I} \Bom) = \mathbf{0}, 
\end{equation}
where $\mathbf{I}$ is the inertia tensor. As is well-known,
this equation requires that $\Bom$ be along a principle
axis of $\mathbf{I}$, and we assume that this is the 
axis having the largest moment of inertia, $I_{3}$. Suppose that the
earth model is deformed
such that its inertia tensor
is perturbed to $\mathbf{I} + \mathbf{i}$.
The steady-state rotation vector will be perturbed to
$\Bom + \bom$, and from the Euler equation we obtain to, first-order accuracy,
\begin{equation}
  \Bom \times \mathbf{j}  +
  I_{3} \bom \times \Bom + \Bom\times (\mathbf{I} \bom) = \mathbf{0},
\end{equation}
where for convenience we have set $\mathbf{j} = \mathbf{i} \Bom$.
Taking the cross product with $\Bom$ and using standard
vector identities, we find that
\begin{equation}
(I_{3}-\mathbf{I}) \bom_{\perp} =   \mathbf{j}_{\perp}, 
\end{equation}
where the subscript $_{\perp}$ is used to denote
projections orthogonal to the reference rotation axes;
note that when restricted to such  vectors the
matrix $I_{3}-\mathbf{I}$ is non-singular.
To fix the component of $\bom$ along the rotation axes,
we note that conservation of angular momentum requires
\begin{equation}
  \|(\mathbf{I}+\mathbf{i})(\Bom+\bom)\|^{2} =   \|\mathbf{I}\Bom\|^{2}, 
\end{equation}
which to first-order yields
\begin{equation}
  - I_{3}  \bom_{\|} =  \mathbf{j}_{\|}, 
\end{equation}
where the subscript $_{\|}$ is used to denote
projections parallel to the reference rotation axes. These results
can be summarised by saying that
\begin{equation}
  \label{eq:torque}
  \tilde{\mathbf{I}} \bom = \mathbf{j},
\end{equation}
with $\tilde{\mathbf{I}}$ a symmetric and invertible matrix.
Relative to a Cartesian basis aligned with the principle axes of $\mathbf{I}$, we note that
\begin{equation}
  \tilde{\mathbf{I}} = \left(
  \begin{array}{ccc}
    I_{3}-I_{1} & 0 & 0 \\
    0 & I_{3}-I_{2} & 0 \\
    0 & 0 & -I_{3}
  \end{array}
  \right),
\end{equation}
where $I_{1}\le I_{2} < I_{3}$ are the principle moments of inertia.
Because $I_{3} \gg I_{3}-I_{1} \ge I_{3}-I_{2}$, it is common
to neglect the component of $\bom$ along the $x_{3}$-axis. This approximation will be made within our  numerical calculations,
but the reciprocity relations  derived below
hold whether or not this is done.

Associated with the equilibrium rotation vector
we can define the centrifugal potential
\begin{equation}
  \Psi = -\frac{1}{2}\|\Bom\times \mathbf{x}\|^{2}, 
\end{equation}
and hence the first-order perturbation, $\psi$, to this potential
is given by
\begin{equation}
  \label{eq:psiid}
  \psi = -(\Bom\times \mathbf{x})\cdot (\bom\times \mathbf{x}).
\end{equation}
The negative gradient of the centrifugal potential gives the centrifugal
acceleration that is needed within the equations of motion,
and so we note the following relations:
\begin{equation}
  \nabla \Psi = \Bom\times(\Bom\times \mathbf{x}), \quad
  \nabla \psi = \Bom \times (\bom \times \mathbf{x}) +
  \bom \times (\Bom \times \mathbf{x}).
\end{equation}

We now need to relate an applied surface load and associated deformation
of the earth model to the perturbed inertia tensor, $\mathbf{i}$.
As a starting point, we note that the earth model's equilibrium
inertia tensor satisfies the identity
\begin{equation}
  \Bom \cdot (\mathbf{I} \Bom) = \int_{M} \rho
  \|\Bom \times \mathbf{x}\|^{2} \dd V, 
\end{equation}
with $\rho$ the equilibrium density, and where $\Bom$
here is an arbitrary vector. When subject to a surface
load, $\sigma$, the first-order perturbation in this quantity
can be written
\begin{equation}
  \Bom \cdot (\mathbf{i} \Bom) = -\int_{M} 2\rho
  \mathbf{u} \cdot [\Bom \times(\Bom\times \mathbf{x})]  \dd V
  + \int_{\partial M} \sigma \|\Bom \times \mathbf{x}\|^{2} \dd S, 
\end{equation}
where standard vector identities have been applied. Here,
the first term is due to internal deformation and the second
is the direct contribution of the load. Changing $\Bom$
within this identity to
$\Bom + \bom'$  and expanding to first-order in $\bom'$ we obtain
\begin{equation}
  \bom' \cdot \mathbf{j} = -\int_{M} \rho \mathbf{u}
  \cdot [\bom' \times (\Bom \times \mathbf{x}) + \Bom \times
    (\bom' \times \mathbf{x})] \dd V
  + \int_{\partial M} \sigma (\Bom\times \mathbf{x})\cdot (\bom'\times \mathbf{x})\dd S, 
\end{equation}
where we recall that $\mathbf{j} = \mathbf{i}\Bom$.
Comparing the  two integrands  with the expression for $\psi$ and $\nabla \psi$
obtained above, we arrive at an identity that will be  useful later
\begin{equation}
  \label{eq:rotid}
  \bom' \cdot \mathbf{j} + \int_{M} \rho \mathbf{u}\cdot \nabla \psi' \dd V
  + \int_{\partial M} \sigma \psi' \dd S = 0.
\end{equation}
Within this equality it should be emphasised that $\psi'$ is the centrifugal potential perturbation
associated with $\bom'$, while $\mathbf{j}$ is the inertia perturbation
associated with the applied load, $\sigma$, and the associated displacement, $\mathbf{u}$.

The derivations within this section have assumed that the earth model is entirely solid.
As shown by \cite{dahlen74},
in considering the static deformation of an earth model with a fluid (outer) core,
the linearised Lagrangian displacement vector is not well-defined
in fluid regions. Instead, Dahlen showed  that  an Eulerian
formulation within  fluid regions could be applied, with all relevant
variables then expressed in terms of the perturbed gravitational
potential. The necessary extensions of our arguments to earth models
containing fluid regions are simple but will not be given because
the form of the reciprocity theorems  is unchanged.

It is also worth commenting that, when calculations
are performed in a spherically symmetric earth model, the
reference inertia tensor, $\mathbf{I}$, is taken to be
that observed for the Earth \citep[c.f.][]{mitrovica2005rotational,mitrovica2011ice}. Here,
there is an assumption that the deformation of the  Earth due to the
applied load can be sufficiently well-approximated by the response
of a spherical earth model. Again, the reciprocity theorems
hold whether or not this approximation is made.

\subsection{Reciprocity theorems including rotational feedbacks}

We start with the equations for a static loading problem
in a rotating earth model in weak form,
\begin{equation}
  \label{eq:eqmr}
  \mathcal{A}(\mathbf{u},\phi\,|\, \mathbf{u}',\phi')
  + \int_{\partial M} (\mathbf{u}'\cdot \nabla \Phi + \phi')\,\sigma \dd S
  + \int_{M} \rho \mathbf{u}'\cdot \nabla \psi \dd V
  + (\tilde{\mathbf{I}}\bom - \mathbf{j})\cdot \bom' = 0.
\end{equation}
Here, $\psi$ is the centrifugal potential associated with the perturbed
rotation vector, $\bom$, and we see that this term is associated with
a tidal force applied to the earth model \citep[c.f.][Appendix B]{bagheri2019tidal}. Note also that
a new test function, $\bom'$, has been added  to enforce the
relation between $\bom$ and the inertia tensor perturbation
$\mathbf{j}$ in eq.(\ref{eq:torque}), the latter term depending implicitly on $\mathbf{u}$
and the applied load, $\sigma$. Following a now familiar procedure,
we let $(\mathbf{u}^{\dagger},\phi^{\dagger},\bom^{\dagger})$
be the solution of the problem subject to a load $\sigma^{\dagger}$, 
and use this to obtain the identity
\begin{equation}
  \int_{\partial M} (\mathbf{u}^{\dagger}\cdot \nabla \Phi + \phi^{\dagger})\,\sigma \dd S
  + \int_{M} \rho \mathbf{u}^{\dagger}\cdot \nabla \psi \dd V
   + \mathbf{j}^{\dagger}\cdot \bom
 =   \int_{\partial M} (\mathbf{u}\cdot \nabla \Phi + \phi)\,\sigma^{\dagger} \dd S
  + \int_{M} \rho \mathbf{u}\cdot \nabla \psi^{\dagger} \dd V
    + \mathbf{j}\cdot \bom^{\dagger}, 
\end{equation}
where we note that terms involving the matrix $\tilde{\mathbf{I}}$ have vanished
due to its symmetry. Using eq.(\ref{eq:rotid}) twice,
this simplifies to 
\begin{equation}
  \label{eq:rec4}
  \int_{\partial M} (\mathbf{u}^{\dagger}\cdot \nabla \Phi + \phi^{\dagger} + \psi^{\dagger})\,\sigma \dd S
 =   \int_{\partial M} (\mathbf{u}\cdot \nabla \Phi + \phi + \psi)\,\sigma^{\dagger} \dd S.
\end{equation}
This is a new result which extends the reciprocity theorem of \cite{trompmitrovica}
to incorporate rotational feedbacks.

To express eq.(\ref{eq:rec4}) in terms of sea level change, we need
merely note that, once rotational feedbacks are present, the relation
between sea level change, $\Delta SL$, and surface deformation is generalised to 
\begin{equation}
  \Delta SL = -\frac{1}{g}(\mathbf{u}\cdot \nabla \Phi + \phi + \psi) + \frac{\Phi_{g}}{g},
\end{equation}
which follows due to the sea surface now lying on an  equipotential of the
combined gravitational and centrifugal potentials. Given this relation,
our earlier argument proceeds  identically to give
\begin{equation}
  \label{eq:rec5}
  \int_{\partial M} \Delta SL^{\dagger}\, \zeta  \dd S
 =   \int_{\partial M} \Delta SL\,\zeta^{\dagger} \dd S,
\end{equation}
which takes the same form as in the non-rotating case in eq.(\ref{eq:rec2}).
From this identity we can readily show the symmetry of the
Green's function and obtain sensitivity kernels for
any functional of the sea level change. These  results are new, extending the
work  of \cite{crawford2018} to include
rotational feedbacks in the case of elastic fingerprint problems.
Extensions of the general viscoelastic adjoint theory to incorporate
rotational feedbacks can be carried out along similar lines, but the results
will be presented elsewhere.

Finally,  if we wish to consider observables other than sea level, it is necessary
to consider the generalised loading problem
\begin{eqnarray}
  \label{eq:eqmrg}
 && \mathcal{A}(\mathbf{u},\phi\,|\, \mathbf{u}',\phi')
  + \int_{\partial M} (\mathbf{u}'\cdot \nabla \Phi + \phi')\,\sigma \dd S
  + \int_{\partial M} (\mathbf{t}\cdot \mathbf{u}' + \zeta_{\phi} \phi') \dd S
   + \int_{M} \rho \mathbf{u}'\cdot \nabla \psi \dd V
  + (\tilde{\mathbf{I}}\bom - \mathbf{j} + \mathbf{k})\cdot \bom'= 0,
\end{eqnarray}
where an additional force term, $\mathbf{k}$, has been included to facilitate observations
depending directly on $\bom$.
Here, we must  account for the contribution of the direct gravitational load, $\zeta_{\phi}$,
to the perturbed inertia tensor, this leading to eq.(\ref{eq:rotid}) being generalised to
\begin{equation}
  \label{eq:rotidg}
  \bom' \cdot \mathbf{j} + \int_{M} \rho \mathbf{u}\cdot \nabla \psi' \dd V
  + \int_{\partial M} (\sigma + \zeta_{\phi}) \psi' \dd S = 0.
\end{equation}
Following the now standard argument, we arrive at the reciprocity theorem
\begin{eqnarray}
  \label{eq:rec6}
&&  \int_{\partial M} \Delta SL^{\dagger} \,\zeta \dd S
  -\frac{1}{g} \int_{\partial M} [\mathbf{t}\cdot \mathbf{u}^{\dagger} + \zeta_{\phi} (\phi^{\dagger}
    + \psi^{\dagger})] \dd S - \frac{1}{g}\mathbf{k}\cdot \bom^{\dagger} \nonumber \\
  &&=  \int_{\partial M} \Delta SL \, \zeta^{\dagger} \dd S
  -\frac{1}{g} \int_{\partial M} [\mathbf{t}^{\dagger}\cdot \mathbf{u} + \zeta^{\dagger}_{\phi} (\phi
  + \psi)] \dd S - \frac{1}{g}\mathbf{k}^{\dagger}\cdot \bom.
\end{eqnarray}
To remove the explicit dependence on the centrifugal potentials, we note from eq.(\ref{eq:psiid}) that
$\psi = -[\mathbf{x}\times(\Bom\times \mathbf{x})]\cdot \bom$ and hence
\begin{eqnarray}
  \label{eq:rec7}
&&  \int_{\partial M} \Delta SL^{\dagger} \,\zeta \dd S
  -\frac{1}{g} \int_{\partial M} (\mathbf{t}\cdot \mathbf{u}^{\dagger} + \zeta_{\phi} \phi^{\dagger}
  ) \dd S - \frac{1}{g}\left[\mathbf{k} - \int_{\partial M}\zeta_{\phi}\, \mathbf{x} \times
    (\Bom\times \mathbf{x}) \dd S\right] \cdot \bom^{\dagger}  \nonumber \\
  && = \int_{\partial M} \Delta SL\,\zeta^{\dagger}  \dd S
  -\frac{1}{g} \int_{\partial M} (\mathbf{t}^{\dagger}\cdot \mathbf{u} + \zeta_{\phi}^{\dagger}  \phi
  ) \dd S - \frac{1}{g}\left[\mathbf{k}^{\dagger}  - \int_{\partial M}\zeta_{\phi}^{\dagger} \, \mathbf{x} \times
    (\Bom\times \mathbf{x}) \dd S\right] \cdot \bom,
\end{eqnarray}
which is an alternative form that can sometimes be useful.

\section{Numerical verification and initial applications}

\begin{figure}
  \centering
   \begin{subfigure}[c]{0.47\textwidth}
     \includegraphics[width=\textwidth]{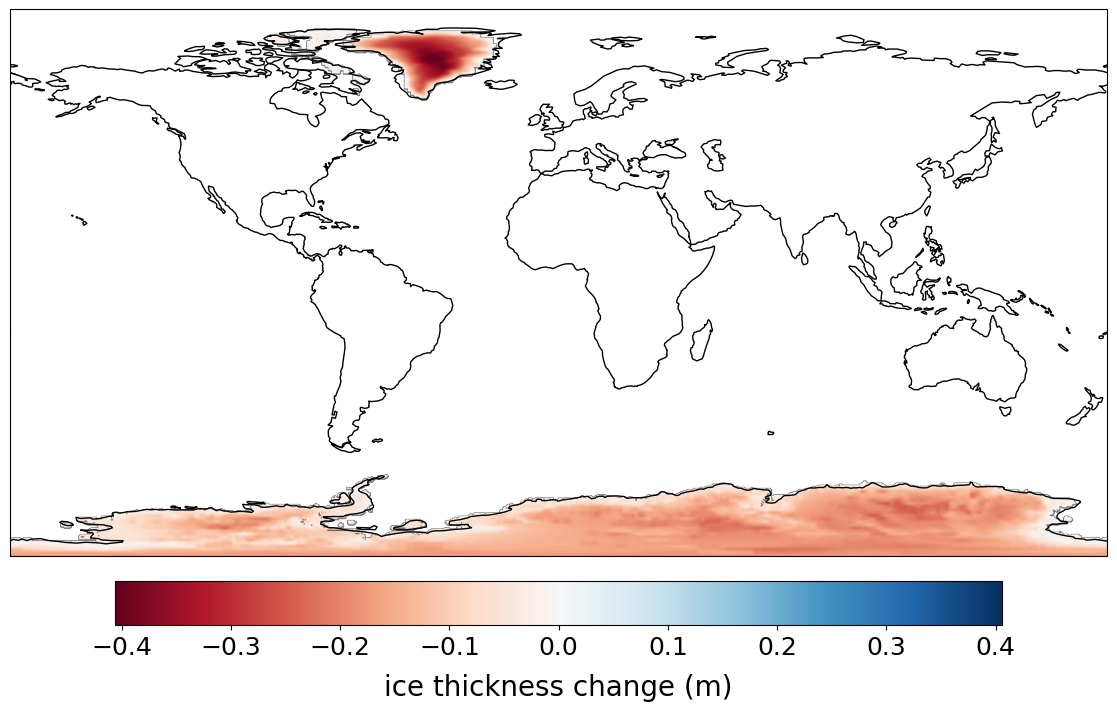}
     \caption{Ice thickness change}
   \end{subfigure} \hfill 
   \begin{subfigure}[c]{0.47\textwidth}
     \includegraphics[width=\textwidth]{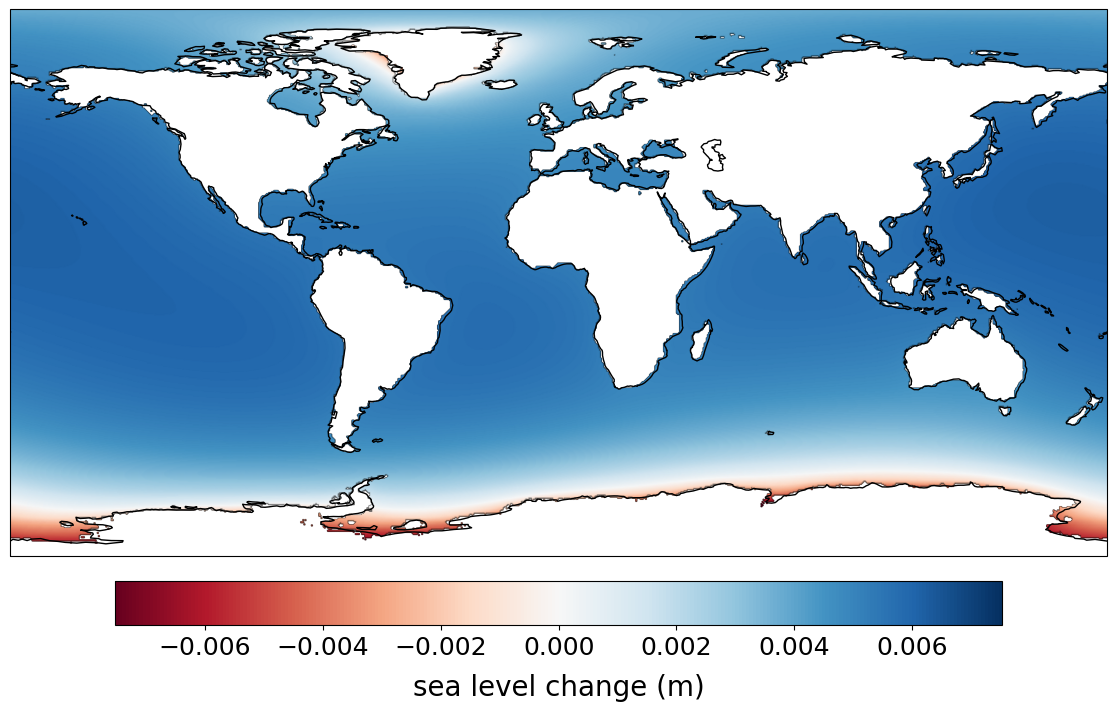}
     \caption{Sea level change}
   \end{subfigure}
   \begin{subfigure}[c]{0.47\textwidth}
     \includegraphics[width=\textwidth]{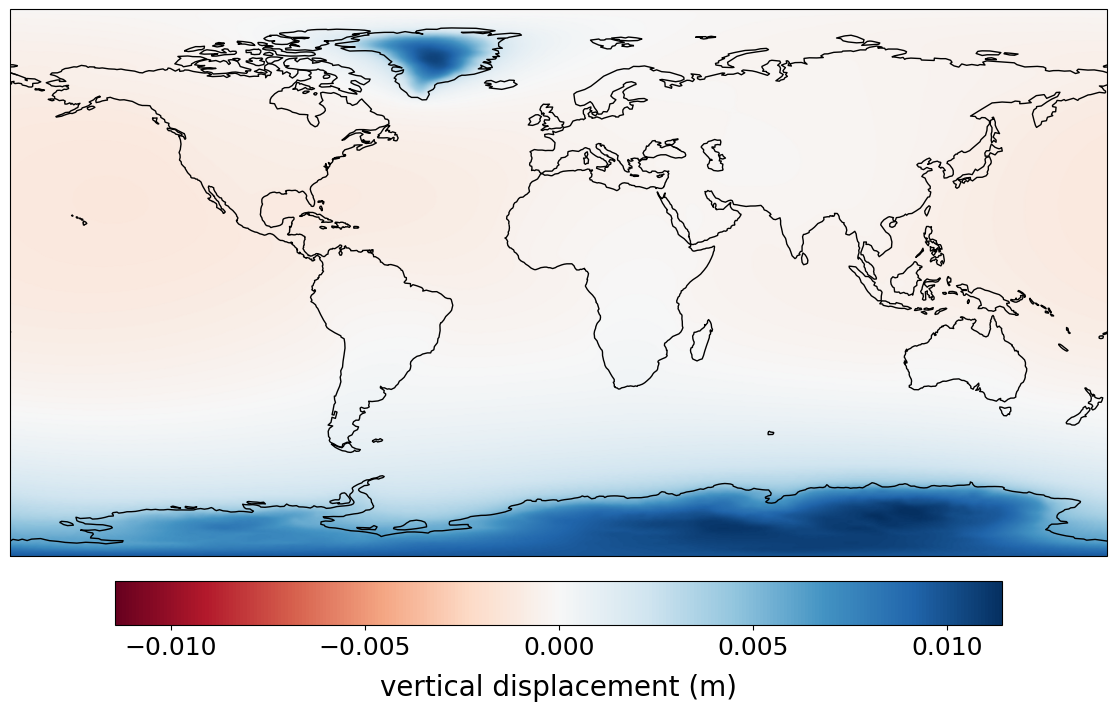}
     \caption{Vertical displacement}
   \end{subfigure} \hfill
   \begin{subfigure}[c]{0.47\textwidth}
     \includegraphics[width=\textwidth]{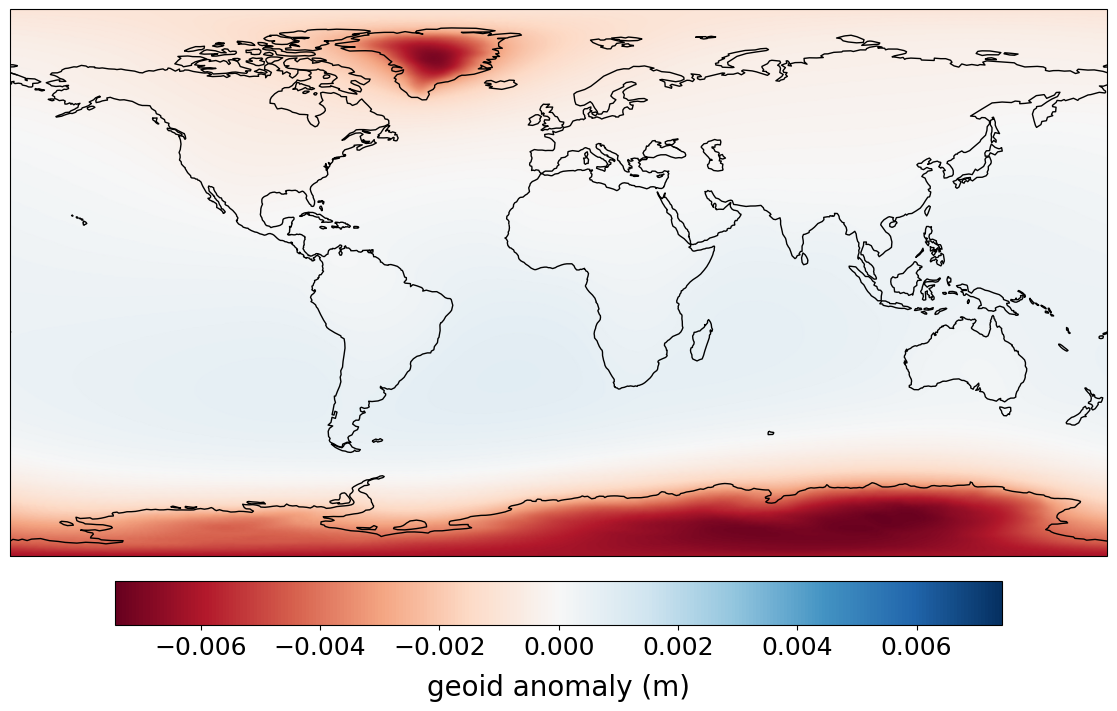}
     \caption{Geoid anomaly}
   \end{subfigure}
   \caption{A summary of the input and results of the forward fingerprint calculation
     used within most of the later numerical examples. In (a) the chosen  ice thickness change is shown.
     In (b) we plot the resulting   sea level change
     obtained through solution of a fingerprint problem. Note that while values of the sea level change are
     only plotted here within the oceans,  this field is defined globally.
     As part of that
     calculation, we also determined the vertical displacement, $\unvec\cdot \mathbf{u}$,
     which is shown in (c),
     and      the gravitational potential perturbation, $\phi$, which is shown in (d) as
     the associated geoid anomaly, $-\phi/g$. Values for the perturbed rotation vector, $\bom$,
   and associated centrifugal potential, $\psi$, were also determined but have not been plotted.}
   \label{fig:forward_calculation}
 \end{figure}

Within this final section, we describe   a range of numerical calculations that
(i) provide  a check on the correctness our theoretical results and (ii) illustrate the potential of these methods within future
practical applications.
All these calculations are done using a spherically symmetric earth model, but we note
that the general theory is applicable within laterally varying elastic earth models. The use of spherically symmetric earth
models is common within fingerprint calculations and it has been shown by \cite{mitrovica2011robustness}
that  lateral variations in elastic structure have only a  minor
effect in this context.

\subsection{Summary of the numerical method}

To solve the elastic fingerprint problem, we apply the standard  pseudo-spectral method  \citep[][]{mitrovica1991postglacial}
with rotational feedbacks incorporated following  \cite{milnemitrovica96,milnemitrovica98}.
 Fast spherical harmonic
transformations are performed using the SHTOOLS library of \cite{wieczorek2018shtools}.
Within all calculations, a truncation degree of $512$ is used, but
it has been verified that truncation at a higher degree does not
appreciably change any of the results.

To account for deformation of the solid earth, we use
loading and tidal Love numbers \citep[][]{love1911some}.
For example, for a surface load, $\sigma$, that has spherical harmonic decomposition
\begin{equation}
  \sigma = \sum_{lm} \sigma_{lm} \,Y_{lm},
\end{equation}
the resulting surface vertical displacement, $u$, and gravitational potential, $\phi$,
have coefficients
\begin{equation}
  u_{lm} = h_{l}\,\sigma_{lm}, \quad \phi_{lm} = k_{l}\, \sigma_{lm},  
\end{equation}
with $h_{l}$ and $k_{l}$ the loading Love numbers (as defined in this work). Here $Y_{lm}$
denotes an  orthonormalised real spherical harmonic of degree $l$
and order $m$ defined following  \cite{wieczorek2018shtools} (note that the definition used for these functions does not
include the Condon-Shortley phase).

The generalised fingerprint problem  involves additional
like-load terms, $\mathbf{t}$ and $\zeta_{\phi}$, introduced within eq.(\ref{eq:eqmnrg}).
Within the following examples, we limit attention to functionals whose dependence on
the surface displacement is only through the vertical component.  It is, therefore, sufficient to assume
\begin{equation}
  \mathbf{t} = \zeta_{u} \nabla \Phi, 
\end{equation}
for a scalar-valued function $\zeta_{u}$.  We can then proceed by introducing suitably generalised loading Love numbers.
If $\zeta_{lm}^{u}$ and $\zeta_{lm}^{\phi}$ are spherical harmonic expansion coefficients
of $\zeta_{u}$ and $\zeta_{\phi}$, the associated coefficients for $u$ and $\phi$ can  be written
\begin{equation}
  u_{lm} =  h_{l}\, \sigma_{lm} + h_{l}^{u} \,\zeta_{lm}^{u} + h_{l}^{\phi} \, \zeta_{lm}^{\phi}, \quad
  \phi_{lm} = k_{l}\,\sigma_{lm} +  k_{l}^{u} \,\zeta_{lm}^{u} + k_{l}^{\phi} \, \zeta_{lm}^{\phi}, 
\end{equation}
where 
$h_{l}^{u}$, $h_{l}^{\phi}$, $k_{l}^{u}$, and $k_{l}^{\phi}$ are the
 generalised loading Love numbers. Calculation of
these Love numbers proceeds in the same manner as the usual,
with the only difference being the form of the forces applied when solving the elastostatic
equations. From these definitions we note the relations
\begin{equation}
  h_{l} = h_{l}^{u} + h_{l}^{\phi}, \quad k_{l} = k_{l}^{u} + k_{l}^{\phi}, 
\end{equation}
while, by applying eq.(\ref{eq:rec3_tmp}), we also have
\begin{equation}
  g\, h_{l}^{\phi} = k_{l}^{u},
\end{equation}
which provides a useful check within numerical calculations.
Values for these generalised loading Love numbers along with the usual tidal Love
numbers have been calculated for PREM \citep[][]{prem} up to degree 4096
using  the radial spectral element method of \cite{alattartromp}.  Translational
non-uniqueness within the degree-one problem is removed by requiring that the combined centre of mass of the earth model
and surface loads is fixed. The remaining rotational degrees of freedom do
not affect  the observables we consider, and hence they   can be ignored.

\subsection{Generating a synthetic data set}

Within Fig.\ref{fig:forward_calculation} we summarise the synthetic data
set  used in later examples. For the purposes of this paper, the specific change in ice thickness
that has been chosen is arbitrary, but for definiteness it was constructed as follows.
Within the northern and southern hemispheres  the changes
in ice thickness were separately set to be  constant fractions of the present-day ice thickness
within ICE6G \citep[][]{arguspeltier,peltier2015space}.
These two contributions were then linearly combined such that  $20$\% of the
ice mass loss was sourced from the northern hemisphere,
while the overall magnitude was fixed by requiring the global mean sea level
change  equal $5$\,mm. Using this change in ice thickness, $\Delta I$, as an input,
the sea level equation was solved to obtain the resulting 
sea level change, $\Delta SL$, vertical displacement, $u$,
gravitational potential perturbation, $\phi$, rotation vector, $\bom$, and
centrifugal potential perturbation, $\psi$, with Fig.\ref{fig:forward_calculation} also showing the first three of
these outputs.

\subsection{Sensitivity kernels for point measurements of sea level}

\begin{figure}
  \centering
   \begin{subfigure}[c]{0.47\textwidth}
     \includegraphics[width=\textwidth]{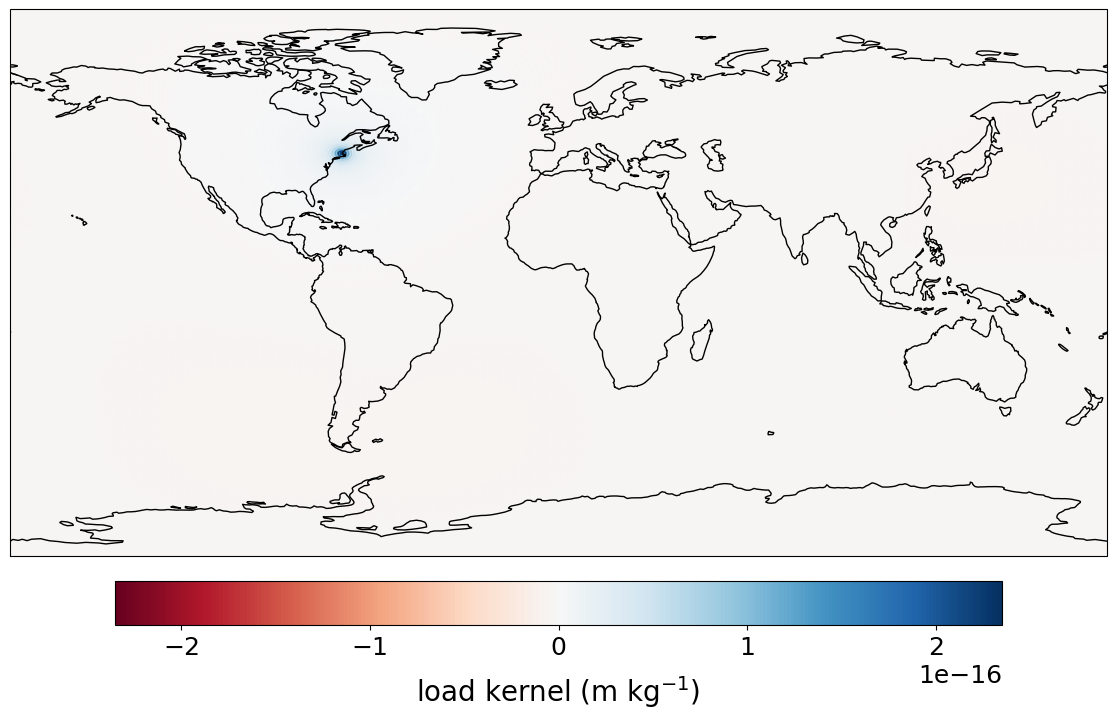}
     \caption{Load kernel}
   \end{subfigure} \hfill
   \begin{subfigure}[c]{0.47\textwidth}
     \includegraphics[width=\textwidth]{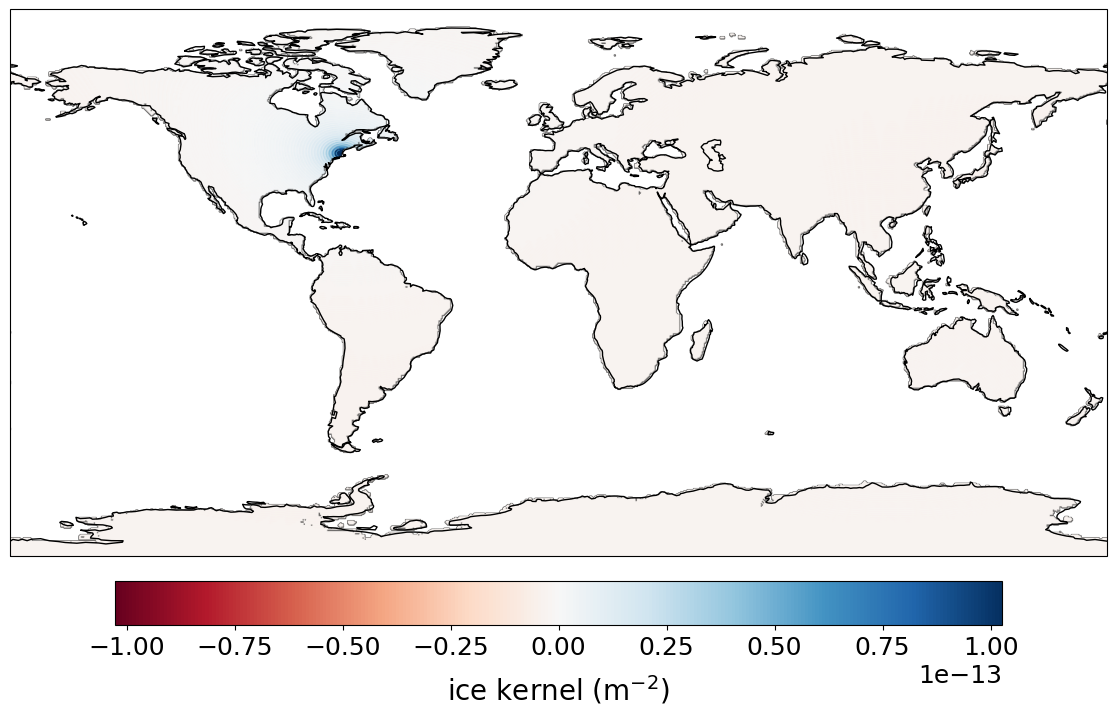}
     \caption{Ice kernel}
   \end{subfigure}
   \begin{subfigure}[c]{0.47\textwidth}
     \includegraphics[width=\textwidth]{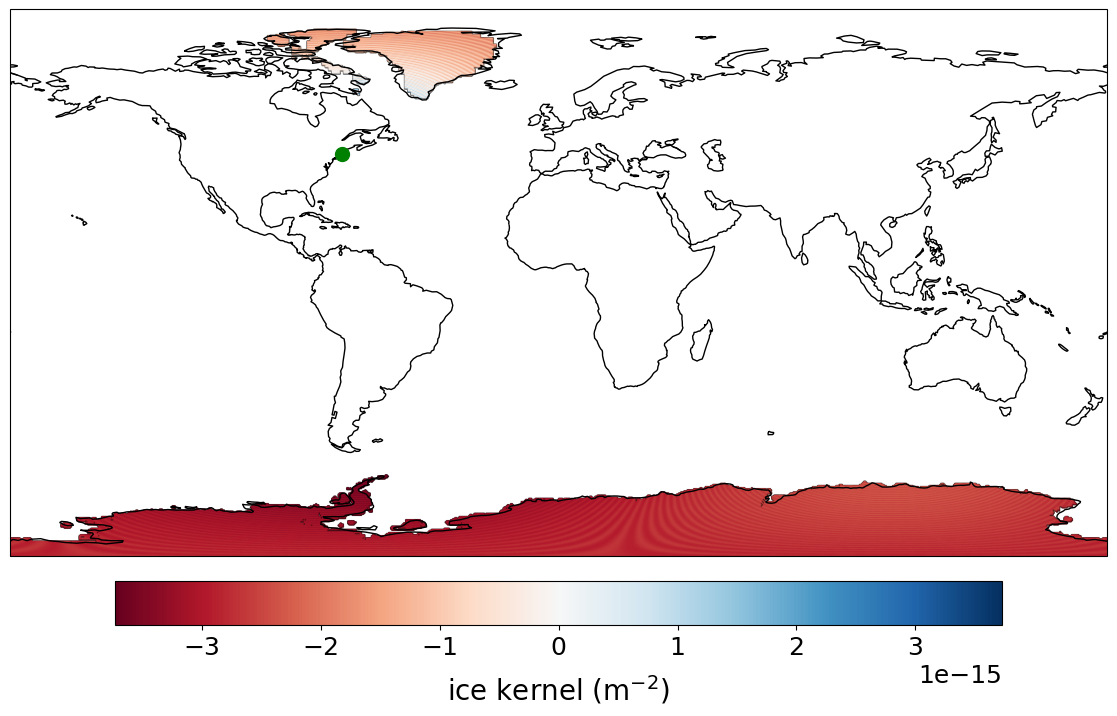}
     \caption{Ice kernel within glaciated regions}
   \end{subfigure} \hfill
   \begin{subfigure}[c]{0.47\textwidth}
     \includegraphics[width=\textwidth]{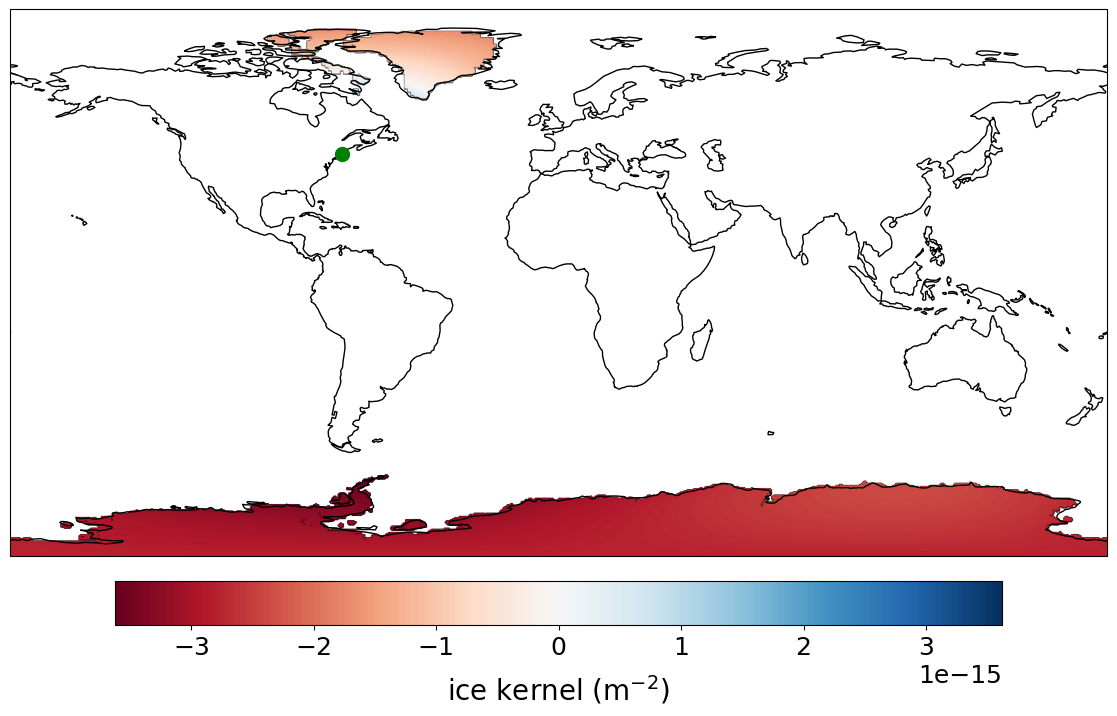}
     \caption{Smoothed ice kernel within glaciated regions}
   \end{subfigure}
   \caption{
     Sensitivity kernels for a point measurement of sea level change in Boston, Massachusetts.
     In (a) we show the kernel with respect to the direct load, $\zeta$, this being
     equal to the adjoint sea level, $\Delta SL^{\dagger}$ obtained by solving the fingerprint
     problem for a point load at Boston. In (b), we show the associated sensitivity kernel
     with respect to ice thickness which takes the form $\rho_{i}\,(1-C)\,\Delta SL^{\dagger}$.
     Note that this kernel is non-zero in all locations where grounded ice \emph{could} be present.
     For most applications, it makes sense to project this kernel onto currently
     glaciated regions, and this is done in (c). Here we can now see more
     structure in the kernel, but also a slight ringing associated with the use of truncated
     spherical harmonic expansions.
     A pragmatic way to avoid this issue is to replace the point
     load with a smooth average over an appropriate length scale. The resulting
     sensitivity kernel is shown in (d) with the measurement corresponding an average
     of the sea level within about one degree of Boston.
     Note that in (c) and (d) the observation point is shown as a green dot. 
  }  
  \label{fig:KSL}
\end{figure}

As noted previously, the sensitivity kernel for a point
measurement of sea level at a location $\mathbf{x}_{0}$ can be obtained
taking the adjoint load, $\zeta^{\dagger}$, to be a delta function at the observation point.
Within Fig.\ref{fig:KSL} we show the results of such a calculation for Boston, Massachusetts.
Within sub-figure (a), the sensitivity kernel, $\Delta SL^{\dagger}$, for this measurement with respect
to the direct load, $\zeta$, is plotted. As might be expected, this sensitivity
is greatest at the observation point and falls in magnitude rapidly away from it.
The value of the kernel at the observation point is positive. This means that
an increase of the direct load in the vicinity of Boston would be associated
with a local rise in sea level. Such behaviour is, of course, consistent with the known physics
of sea level change.

To check quantitatively whether this kernel is correct, we can numerically evaluate
and compare either side of eq.(\ref{eq:rec5})  using the load and sea level
shown in Fig.\ref{fig:forward_calculation}. For the left hand side of the
identity, we evaluate $\int_{\partial M} \Delta SL^{\dagger}\,\zeta \dd S$ numerically
using the solution, $\Delta SL^{\dagger}$, of the adjoint  problem along with the direct load, $\zeta$, applied
within
the forward  problem. The right hand side of the identity is given by $\int_{\partial M} \Delta SL\,\zeta^{\dagger} \dd S$, and this can
be determined by numerical integration or simply through evaluation
of $\Delta SL$ at the observation point. The relative difference between the values
obtained was of order $10^{-7}$ which is about the  accuracy of the numerical calculations (no formal error analysis of the numerical methods
has been undertaken). This test has been repeated using a range of different direct loads, while similar comparisons have been
made for each   the kernels corresponding to each of the  observables discussed below.

The sensitivity kernel of sea level with respect to ice thickness is given by $\rho_{i}\,(1-C)\, \Delta SL^{\dagger}$
which is plotted in Fig.\ref{fig:KSL} (b) for the Boston measurement. Here
it is worth noting that this kernel can be non-zero anywhere on land.
It makes sense, however, in most applications to
multiply this kernel by a function that equals one in glaciated regions  and
by zero elsewhere.  This has been done in  Fig.\ref{fig:forward_calculation} (c),
and now we are able to see more interesting spatial variability. This kernel
does, however, contain some high-wavenumber oscillations, with this feature being more pronounced
when lower truncation degrees are used within the calculations. The reason is that
a delta function cannot be accurately represented using truncated spherical
harmonic expansions (nor in any finite-dimensional basis). For some applications
these oscillations have no practical effect and  can be ignored. However, a simple and
pragmatic way to remove the oscillations is to replace the point load by
an appropriate smooth averaging function \citep[c.f.][Appendix E]{alattartromp}.
The resulting kernel is shown in Fig.\ref{fig:forward_calculation} (d)
obtained using an averaging function with a characteristic length-scale
of one degree (in more detail, the Green's function for the heat
equation on a sphere was used, with the time chosen such that
diffusion over the desired length-scale has occurred). The result
of this latter calculation can be compared with Fig.1 in
\cite{mitrovica2018quantifying} which shows a sensitivity kernel
for sea level at Boston (along with several other locations) to variations in ice
thickness over Greenland. Within that paper, a 
finite-difference approach was used to obtain these sensitivity
kernels, and this required many forward calculations.  Using our present methods such
kernels can be determined for any desired location through a single fingerprint calculation.

\subsection{Sensitivity kernels for vertical displacement measurements}

\begin{figure}
  \centering
  \begin{subfigure}[c]{0.47\textwidth}
     \includegraphics[width=\textwidth]{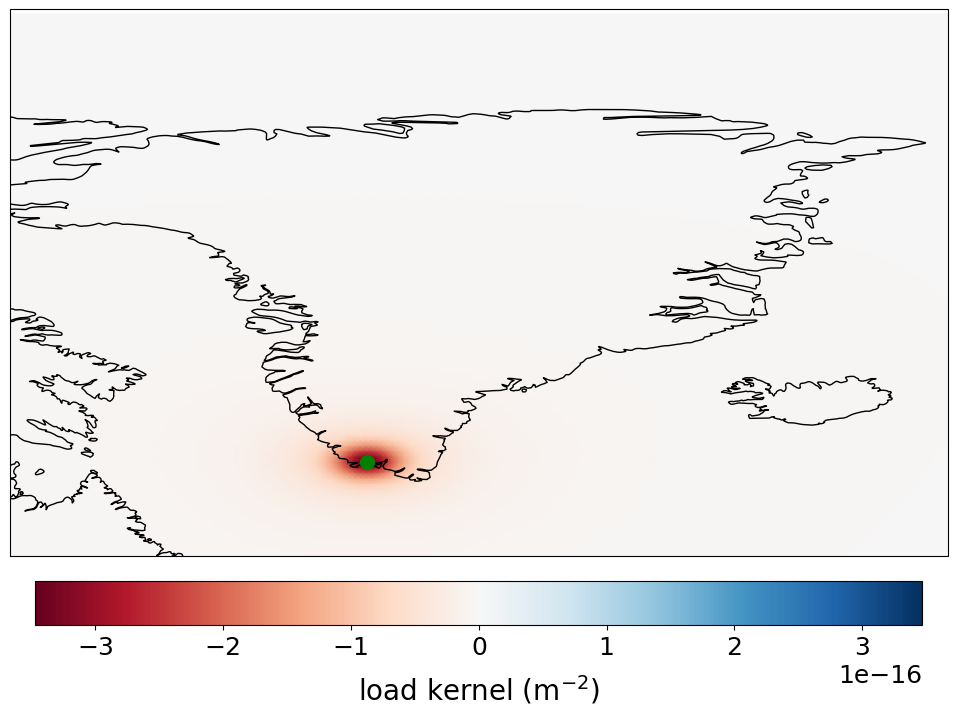}
     \caption{Load kernel for SENU}
   \end{subfigure} \hfill
   \begin{subfigure}[c]{0.47\textwidth}
     \includegraphics[width=\textwidth]{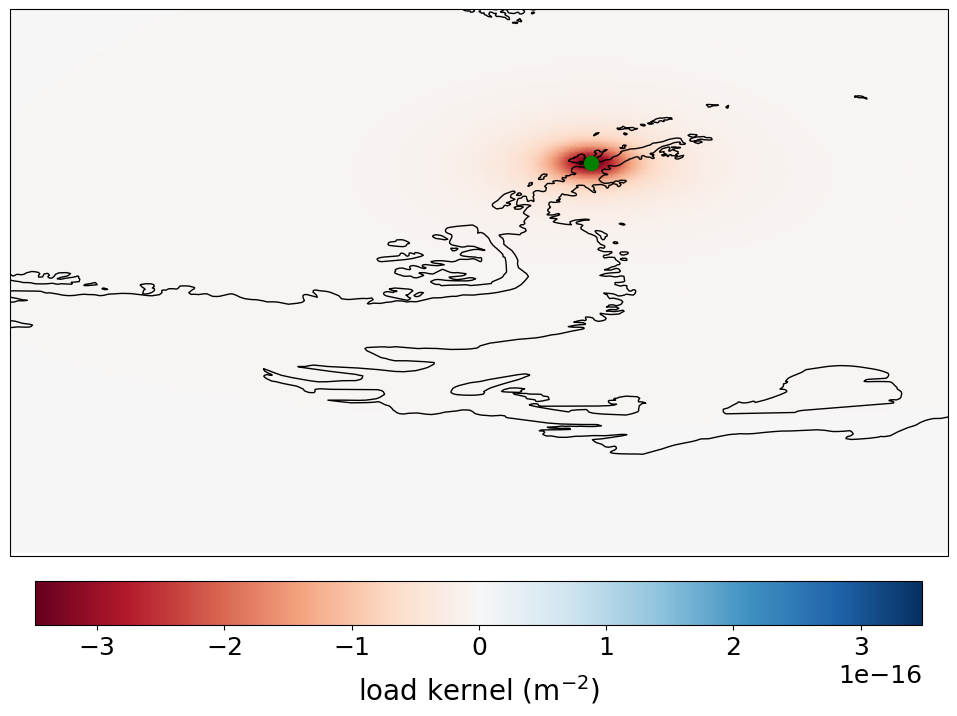}
     \caption{Load kernel for DUPT}
   \end{subfigure}
  \caption{
    The sensitivity kernel of vertical displacement at the GPS stations (a) SENU
    in Greenland and  (b) DUPT in Antarctica  with respect to direct load.
    As before, the point load occurring within the generalised fingerprint
    problem has been smoothed over a distance of one degree to
    remove high-wavenumber oscillations from the kernel. In each plot,
    the observation point is indicated by a green dot.
  }  
  \label{fig:Ku}
\end{figure}

Consider a measurement of vertical displacement at a surface location $\mathbf{x}_{0}$,  made using
the Global Positioning System (GPS) \citep[e.g.][]{khan2010spread}. We can determine the sensitivity
kernel for such a measurement with respect to the direct load by solving the generalised fingerprint problem
subject to the following adjoint loads:
\begin{equation}
\zeta^{\dagger} = 0, \quad \zeta_{u}^{\dagger}(\mathbf{x})  = -\delta(\mathbf{x},\mathbf{x}_{0}), 
\quad \zeta_{\phi}^{\dagger} = 0, \quad \mathbf{k}^{\dagger} = \mathbf{0}.
\end{equation}
The result of two such calculations is shown in Fig.\ref{fig:Ku} for
SENU station in Greenland and DUPT in Antarctica.  Here, as before, we  
have used smoothed point loads to remove non-physical oscillations from the kernel. It should be emphasised that in these examples,
the displacements are computed relative to the combined centre of mass of earth model and surface loads. A translation of
these results to any other reference system (e.g., the centre of mass of the solid Earth) is straightforward.
The aim of these examples is merely to show that the calculation of such sensitivity kernels can be readily done,
with the practical application of these ideas  to be discussed  in future work.

\subsection{Sensitivity kernels for gravitational potential coefficients}

\begin{figure}
  \centering
  \begin{subfigure}[c]{0.47\textwidth}
     \includegraphics[width=\textwidth]{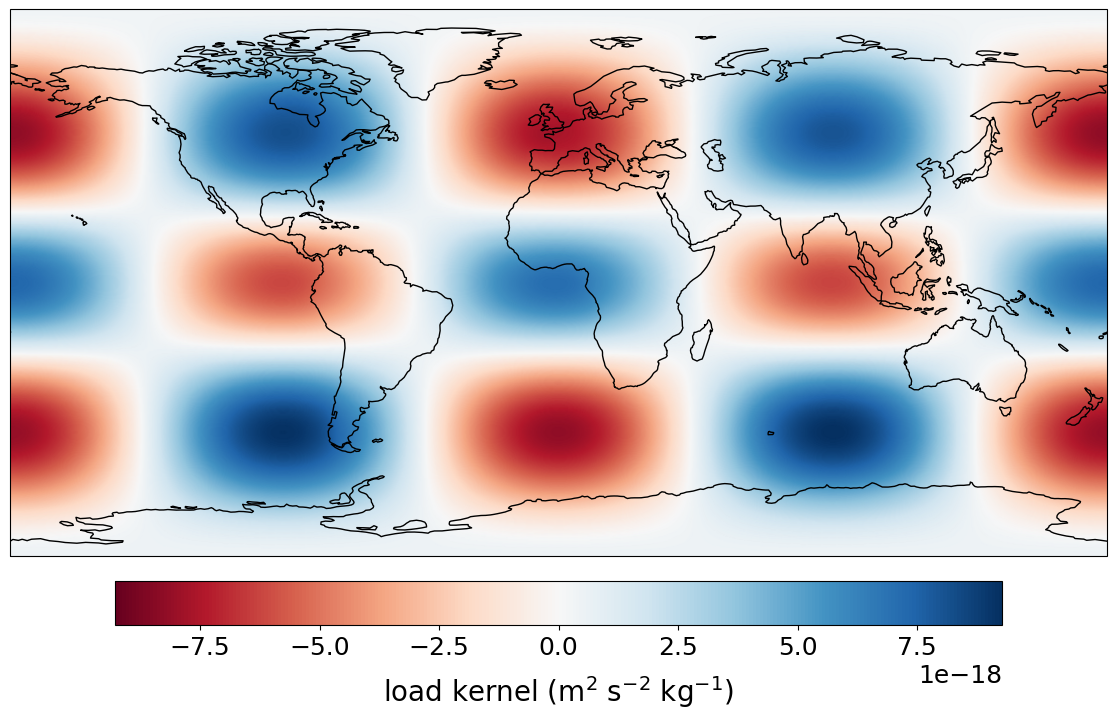}
     \caption{Load kernel}
   \end{subfigure} \hfill
   \begin{subfigure}[c]{0.47\textwidth}
     \includegraphics[width=\textwidth]{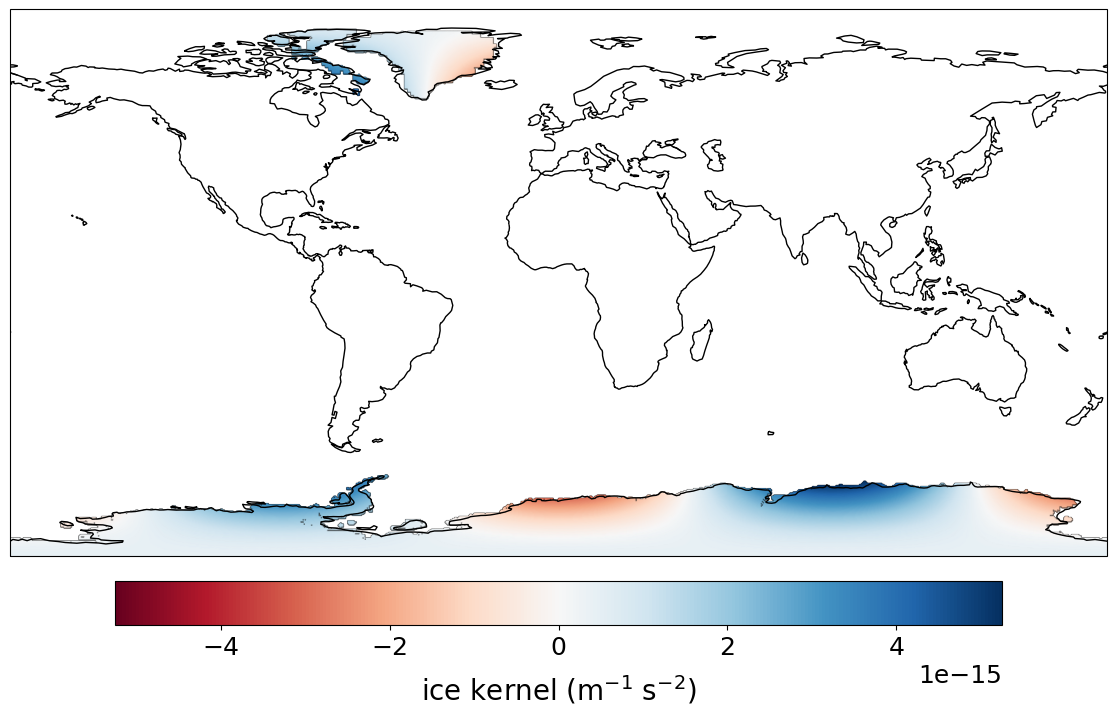}
     \caption{Ice kernel}
   \end{subfigure}
  \begin{subfigure}[c]{0.47\textwidth}
     \includegraphics[width=\textwidth]{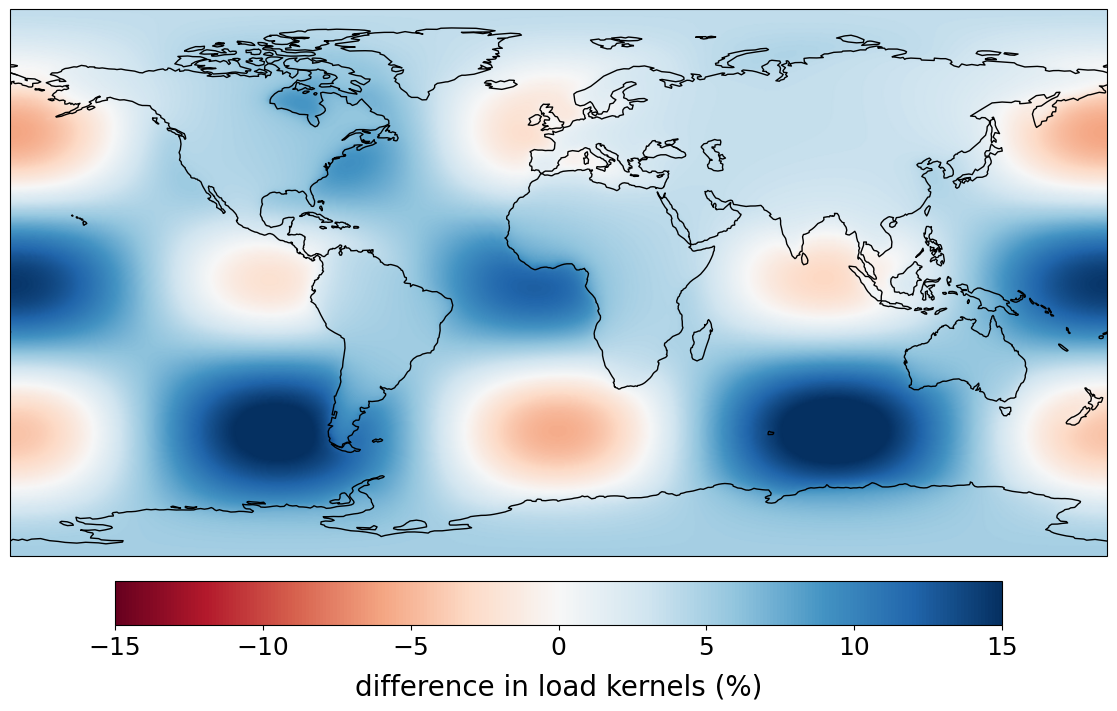}
     \caption{Difference with  load kernel that neglects induced water loading}
   \end{subfigure} \hfill
   \begin{subfigure}[c]{0.47\textwidth}
     \includegraphics[width=\textwidth]{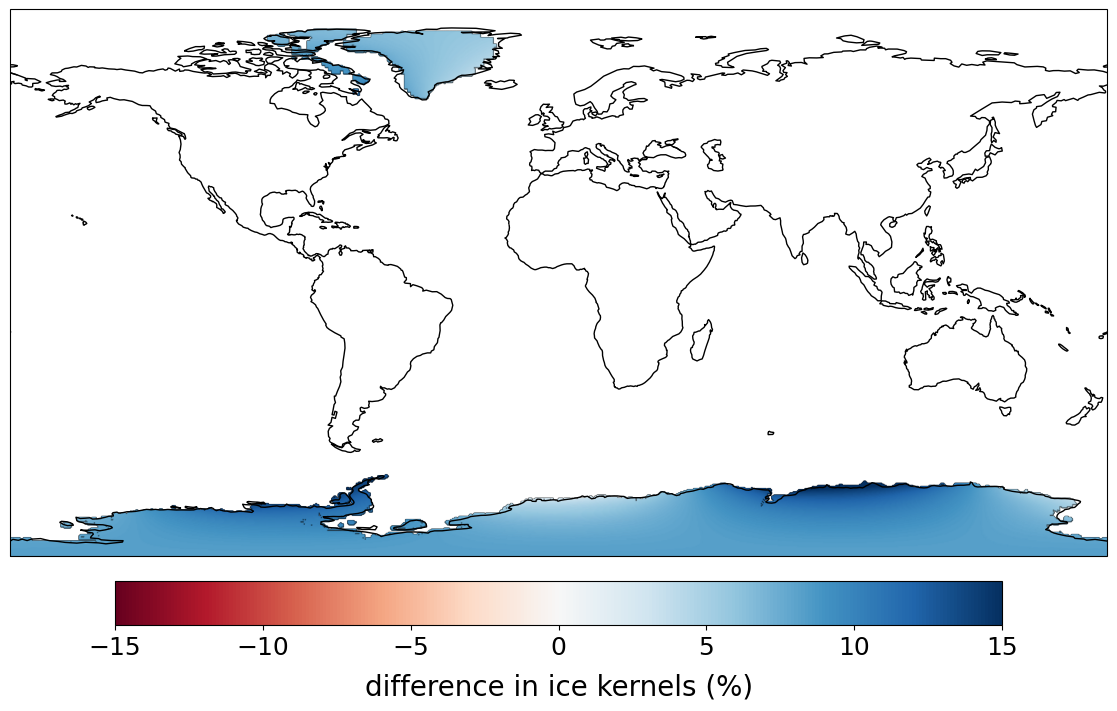}
     \caption{Difference with ice kernel that neglects induced water loading}
   \end{subfigure}
   \caption{
     In (a) we show the sensitivity kernel for the spherical harmonic coefficient,
     $\phi_{4,2}$, of the gravitational potential perturbation to the direct load,
     $\zeta$. The kernel for the same measurement but with respect to
     ice thickness over glaciated regions is then
     plotted in (b). Within the calculation of these sensitivity kernels,
     the effect of both the direct load and a gravitationally self-consistent
     water load have been taken into account. Their form is compared in (c) and (d)
     to simpler kernels that account only for the  direct load. In (c)
     we plot the difference between the kernel in (a) and this simpler kernel
     defined with respect to $\zeta$. Values are shown as a percentage
     defined relative to the maximum absolute value of the kernel in (a).
     In (d) we show the analogous difference for the kernel with respect to
     ice thickness over glaciated regions.
  }  
  \label{fig:Kphi}
\end{figure}

We consider the measurement of a spherical harmonic coefficient of
the gravitational potential at degree $l\ge 2$ as might be obtained
using GRACE or GRACE-FO \citep[e.g][]{tapley2004gravity,landerer2020extending}.
For definiteness, we use real, fully-normalised spherical harmonics,
denoting the $(l,m)$th function by $Y_{lm}$. In determining the adjoint loads
for this application,  we use eq.(\ref{eq:rec7}) to select
a value for $\mathbf{k}^{\dagger}$ such that the
dependence on $\bom$ is removed from the right hand side, with this result:
\begin{equation}
\zeta^{\dagger} = 0, \quad \zeta_{u}^{\dagger}(\mathbf{x})  = 0, 
\quad \zeta_{\phi}^{\dagger} = -g \,Y_{lm}, \quad \mathbf{k}^{\dagger} = \int_{\partial M}\zeta_{\phi}^{\dagger} \, \mathbf{x} \times
    (\Bom \times \mathbf{x}) \dd S.
\end{equation}
Example sensitivity kernels for $l = 4$ and $m = 2$ are
shown in Fig.\ref{fig:Kphi}. The kernel with respect to
$\zeta$ is plotted globally in (a), while that for ice thickness
is projected onto glaciated regions in (b).

\begin{figure}
  \centering
  \begin{subfigure}[c]{0.47\textwidth}
     \includegraphics[width=\textwidth]{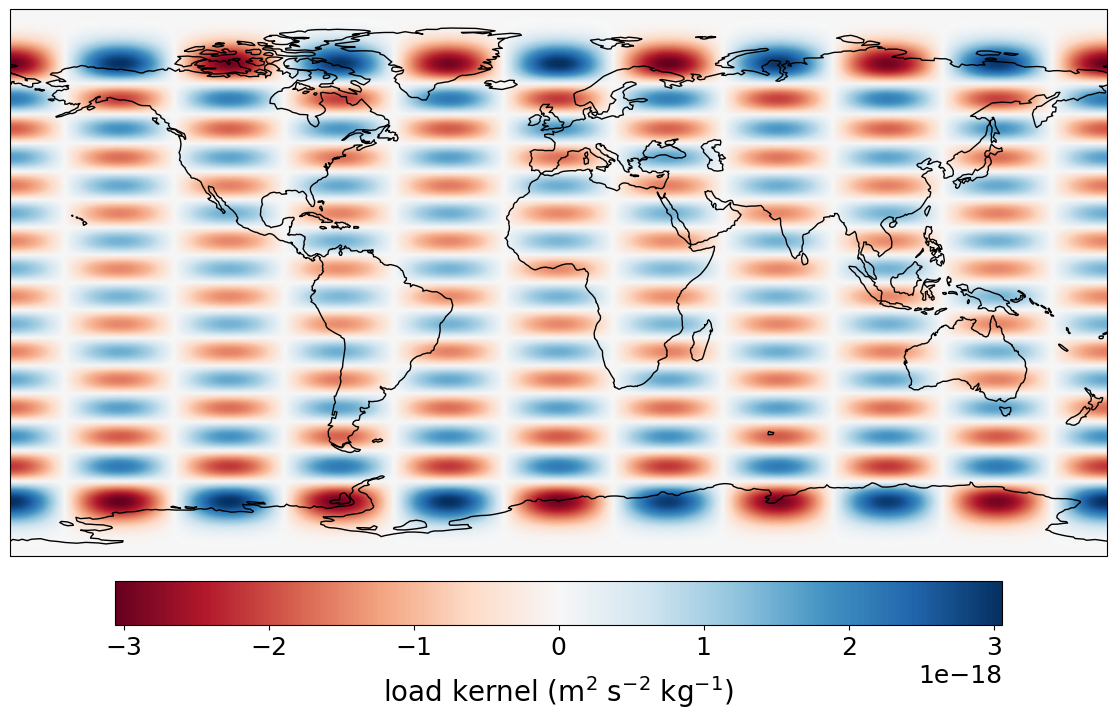}
     \caption{Load kernel}
   \end{subfigure} \hfill
   \begin{subfigure}[c]{0.47\textwidth}
     \includegraphics[width=\textwidth]{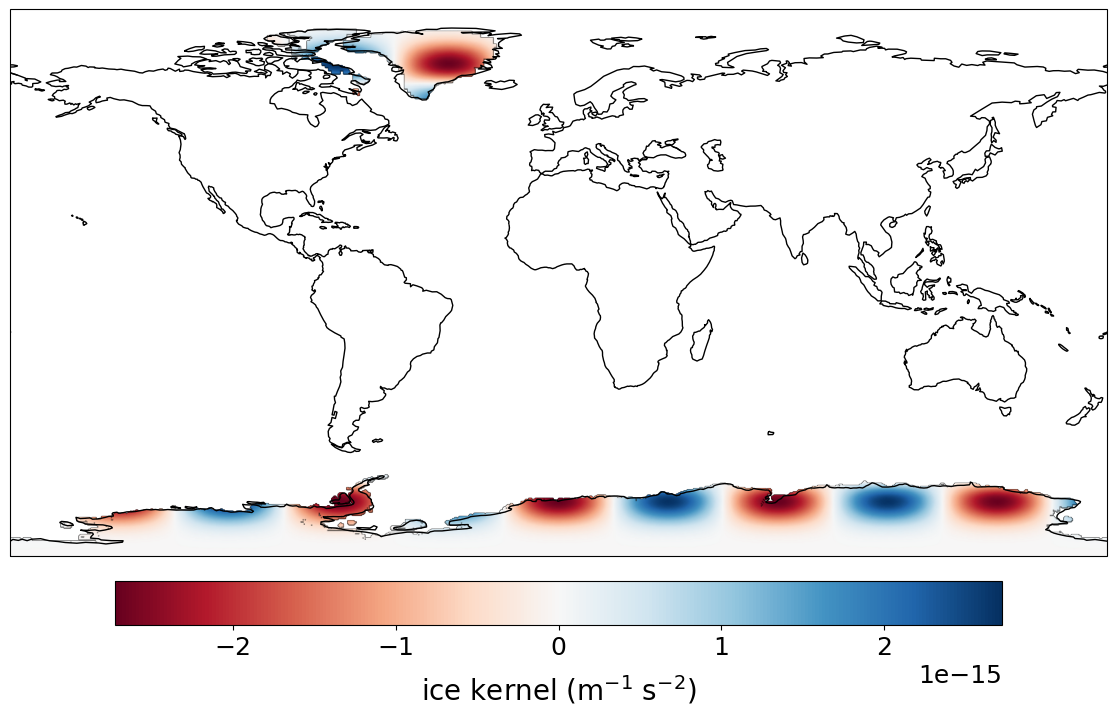}
     \caption{Ice kernel}
   \end{subfigure}
  \begin{subfigure}[c]{0.47\textwidth}
     \includegraphics[width=\textwidth]{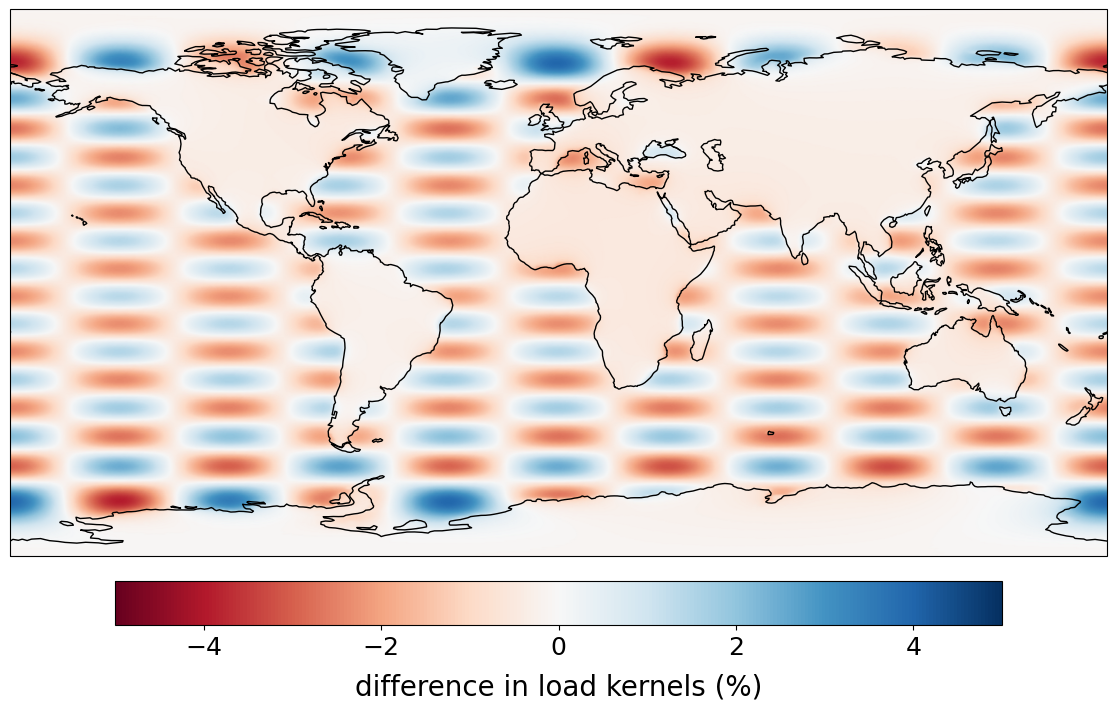}
     \caption{Difference with  load kernel that neglects induced water loading}
   \end{subfigure} \hfill
   \begin{subfigure}[c]{0.47\textwidth}
     \includegraphics[width=\textwidth]{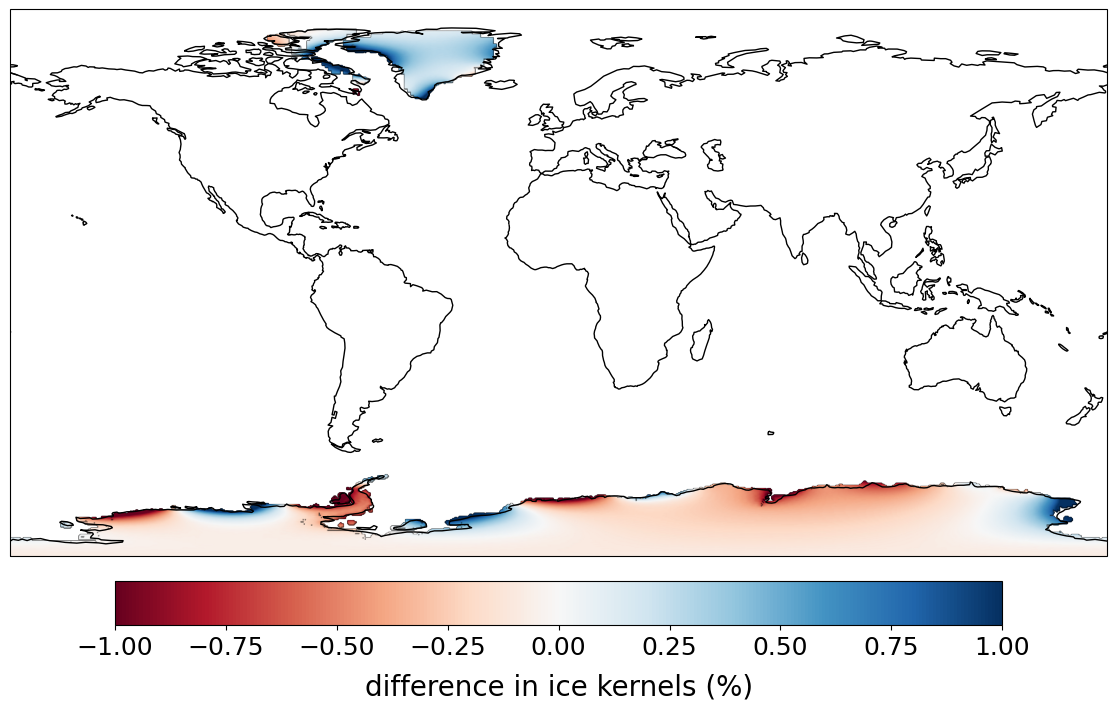}
     \caption{Difference with ice kernel that neglects induced water loading}
   \end{subfigure}
   \caption{
     As in Fig.\ref{fig:Kphi}, but showing results for the $l = 20$, $m = 5$ coefficient
     of the gravitational potential.
  }  
  \label{fig:Kphi2}
\end{figure}

The kernels just shown reflect  contributions to the potential coefficient associated
with both the direct load and the gravitationally
self-consistent water load it must induce. They can be
usefully contrasted with kernels that account
only for the direct load. In the latter case
we can  simply write
\begin{equation}
  \phi_{lm} = k_{l}\,\zeta_{lm}, 
\end{equation}
with $k_{l}$ the loading Love number for $\phi$ at degree $l$. Given this relation, we then have
\begin{equation}
  \phi_{lm} = \frac{k_{l}}{b^{2}}\int_{\partial M} \zeta \, Y_{lm}\dd S, 
\end{equation}
with $b$ the radius of the earth model, and hence we conclude that the relevant kernel is $\frac{k_{l}}{b^{2}}Y_{lm}$.
In Fig.\ref{fig:Kphi} (c),
we show the relative difference between the sensitivity kernel with respect to the
direct load in (a)  and the simpler kernel  just defined. As might be
expected, the relative differences are largest within the oceans, but
even on land they can be of the order of several percent. Finally, in (d) we 
plot  the corresponding difference between the ice kernels
over glaciated regions, observing that its magnitude
can in  places be upwards of ten percent. Within Fig.\ref{fig:Kphi2}
we show  the same thing  for a higher-degree coefficient
of the gravitational potential, specifically $l = 20$ and $m = 5$.
Here, we see smaller differences due to the neglect of induced water loads,
but again the differences are most significant within the oceans.
These results are consistent with the work of \cite{sterenborg2013bias} who  showed
that  methods for analysing GRACE or GRACE-FO data that do not account for induced water
loads suffer from systematic biases. Within the present examples we see, in particular, that the importance
of the induced water load is scale-dependent, decreasing as the dominant wave-number
of the direct load increases.

\begin{figure}
  \centering
  \begin{subfigure}[c]{0.47\textwidth}
     \includegraphics[width=\textwidth]{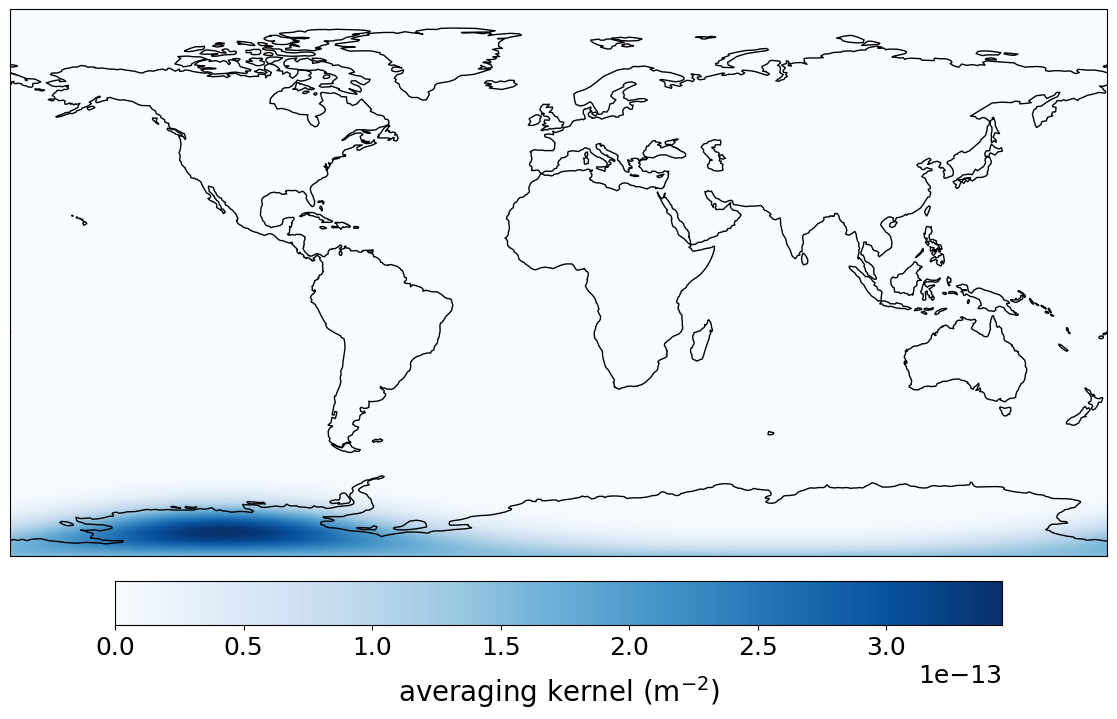}
     \caption{West Antarctica averaging function}
   \end{subfigure} \hfill
   \begin{subfigure}[c]{0.47\textwidth}
     \includegraphics[width=\textwidth]{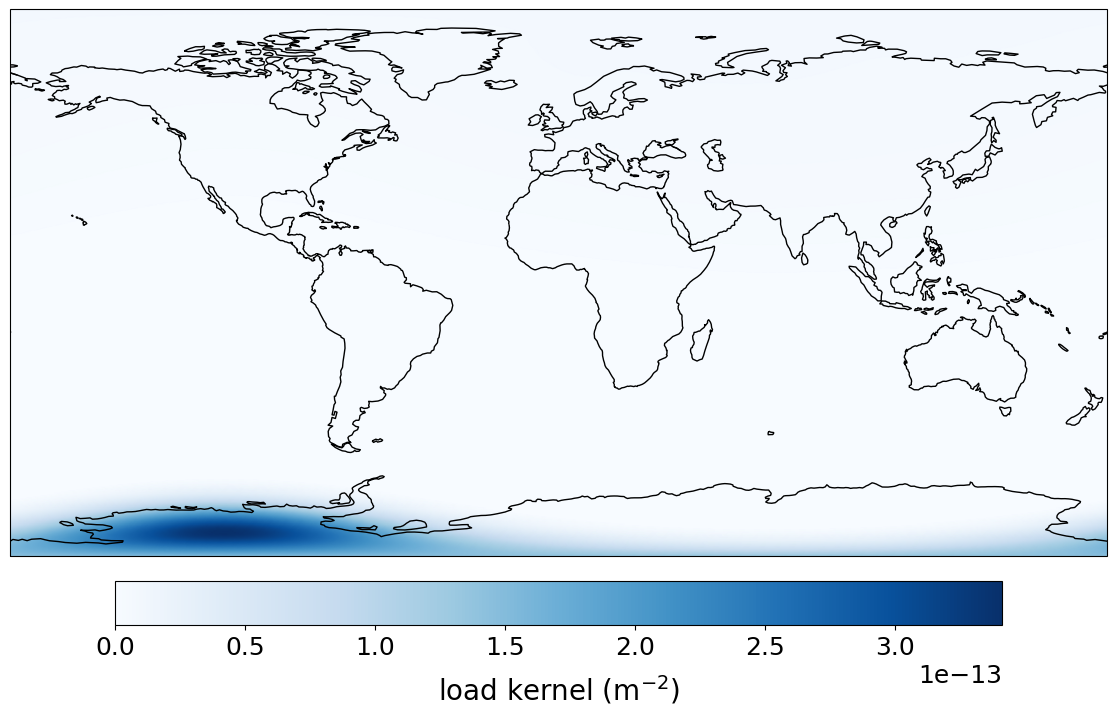}
     \caption{Load sensitivity kernel}
\end{subfigure}
  \begin{subfigure}[c]{0.47\textwidth}
     \includegraphics[width=\textwidth]{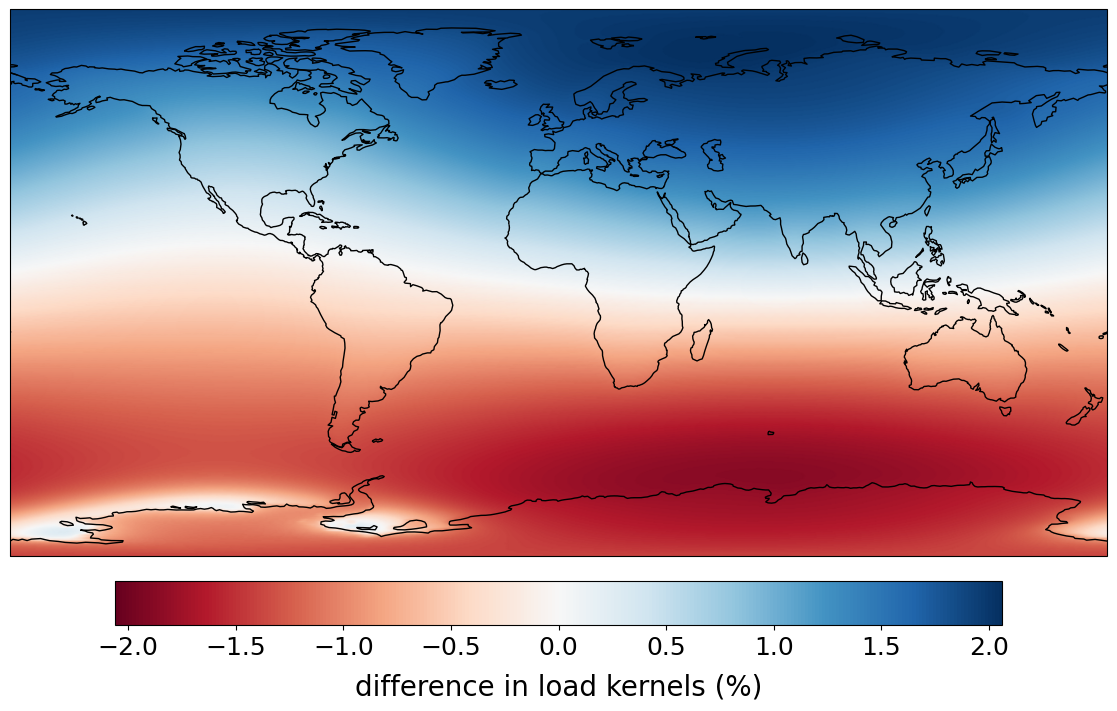}
     \caption{Difference between load kernel and averaging function}
   \end{subfigure} \hfill
   \begin{subfigure}[c]{0.47\textwidth}
     \includegraphics[width=\textwidth]{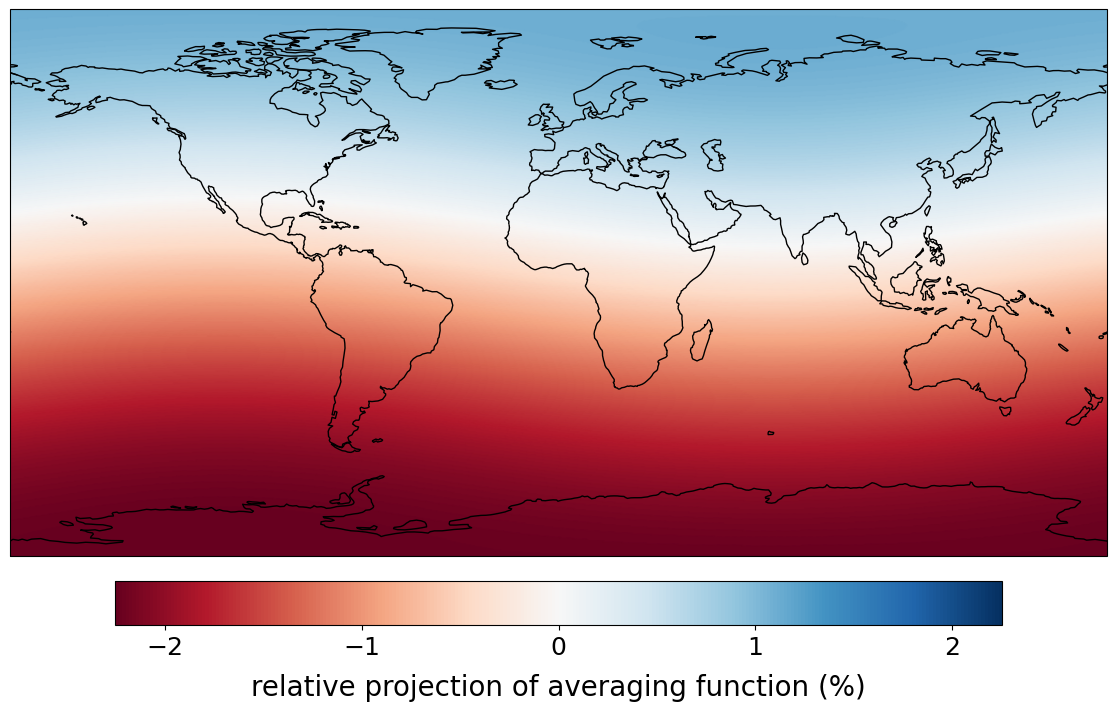}
     \caption{Relative degree-one projection of averaging function}
   \end{subfigure}
   \caption{
     In (a) we show a Gaussian load averaging function targeted at West Antarctica
     with a half-width of 800\,km. In (b) we plot the resulting sensitivity kernel
     with respect to the direct load. The relative difference between the
     kernel and the averaging function is shown in (c) with the values
     normalised by the maximum value of the latter field. In (d)
     we show $\bar{w}-w$ there $\bar{w}$ is the projection
     of the averaging function onto degrees $l \ge 2$.
  }  
  \label{fig:GRACE_average_antarctica}
\end{figure}

\begin{figure}
  \centering
  \begin{subfigure}[c]{0.47\textwidth}
     \includegraphics[width=\textwidth]{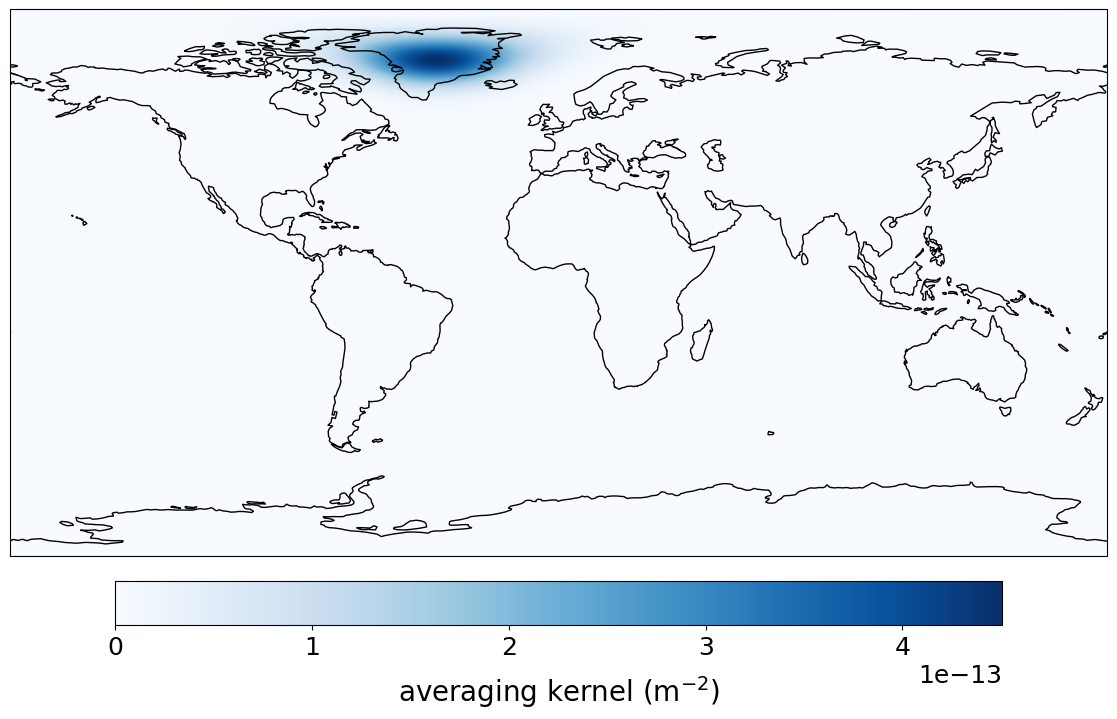}
     \caption{West Greenland averaging function}
   \end{subfigure} \hfill
   \begin{subfigure}[c]{0.47\textwidth}
     \includegraphics[width=\textwidth]{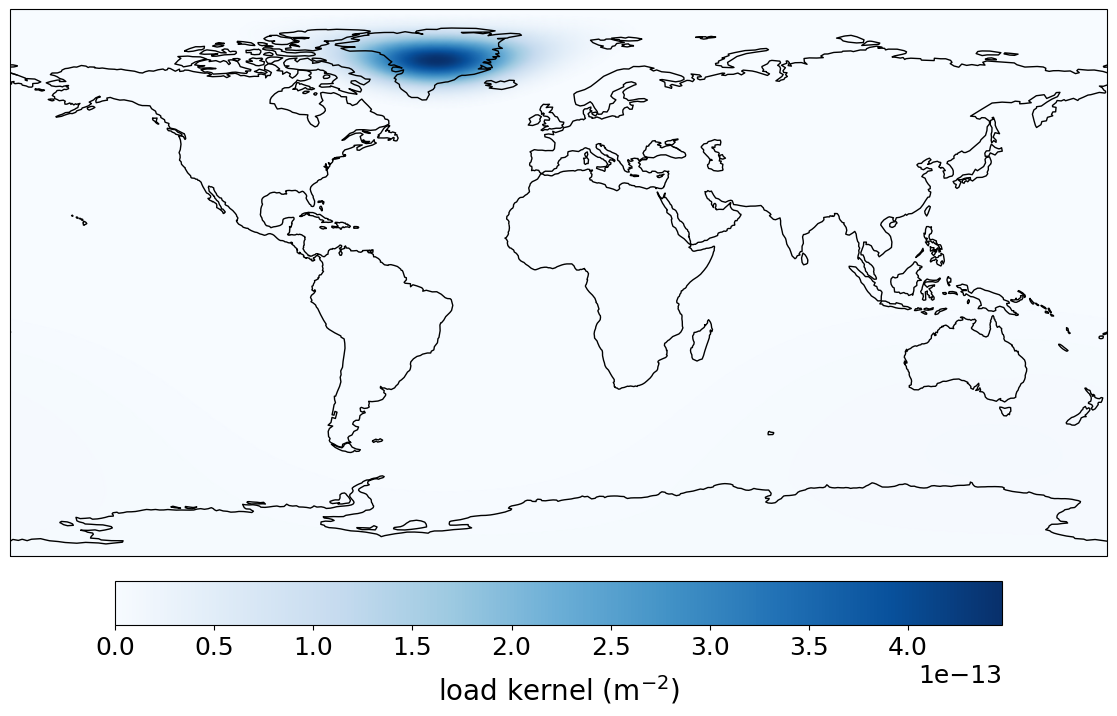}
     \caption{Load sensitivity kernel}
\end{subfigure}
  \begin{subfigure}[c]{0.47\textwidth}
     \includegraphics[width=\textwidth]{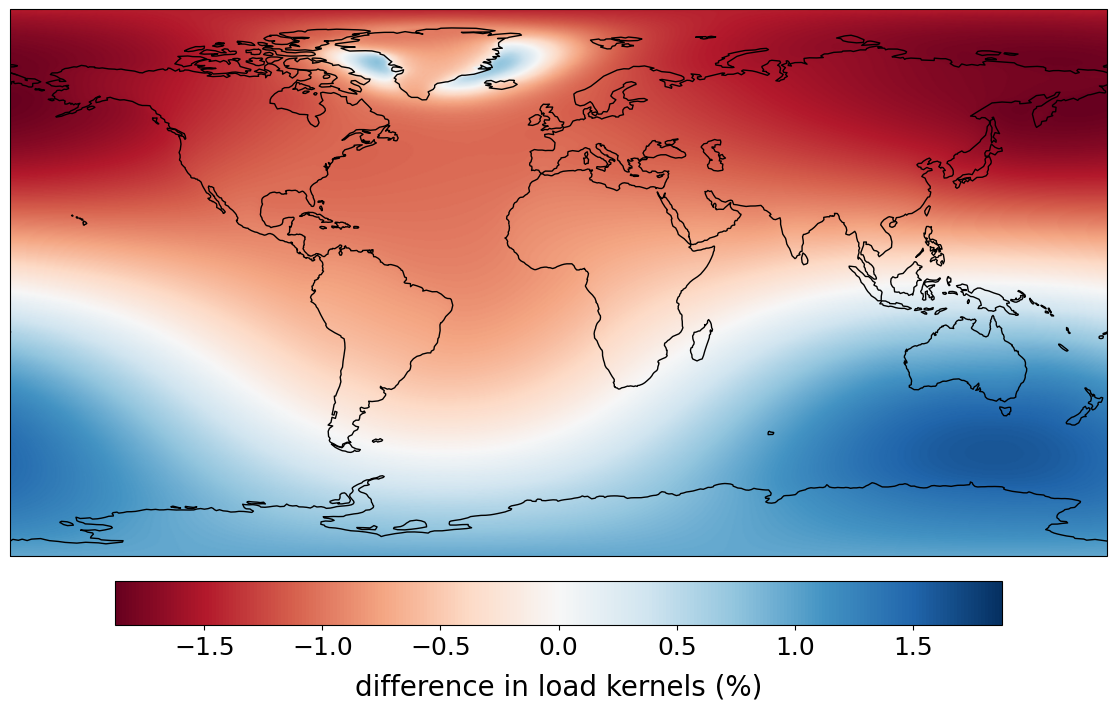}
     \caption{Difference between load kernel and averaging function}
   \end{subfigure} \hfill
   \begin{subfigure}[c]{0.47\textwidth}
     \includegraphics[width=\textwidth]{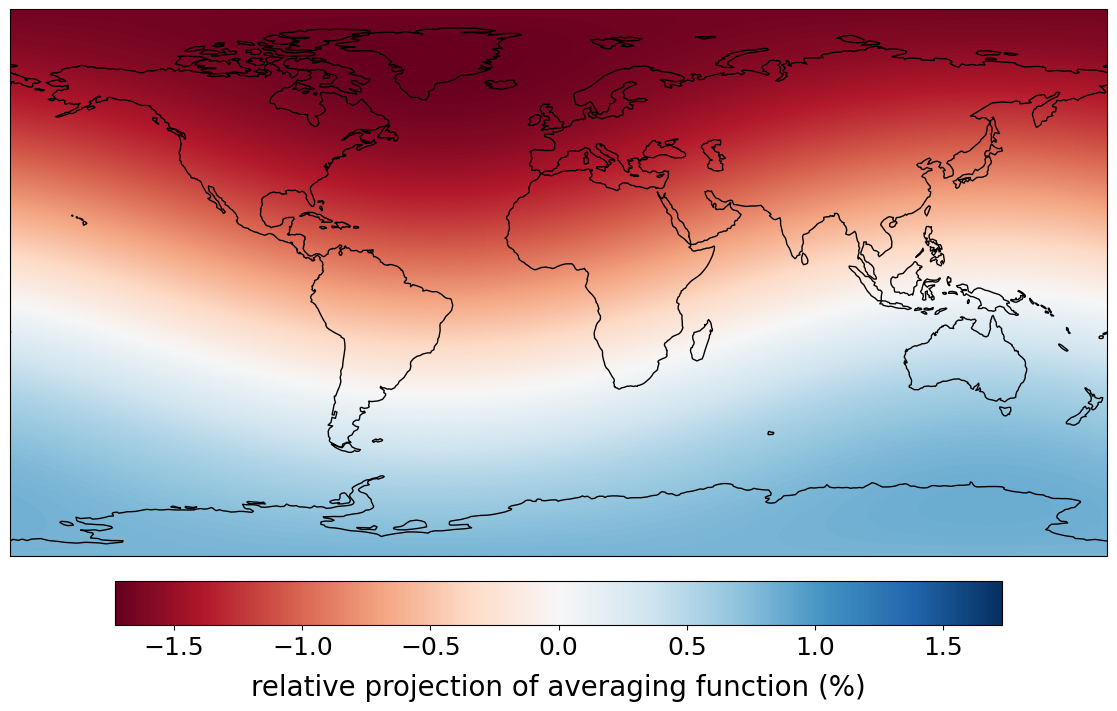}
     \caption{Relative degree-one projection of averaging function}
   \end{subfigure}
   \caption{
     As for Fig.\ref{fig:GRACE_average_antarctica} but for an averaging function
     targeted at Greenland with a half-width of $700$\,km.
  }  
  \label{fig:GRACE_average_greenland}
\end{figure}

Building on the previous example, we now consider the use of GRACE
coefficients to estimate surface mass changes within ice sheets \citep[e.g.][]{velicogna2006acceleration}
following the
widely used method of \cite{wahr1998time} and \cite{swenson2002methods}.
In the notations of this present paper, the approach starts with relation
\begin{equation}
  \phi_{lm} = k_{l}\,\sigma_{lm}, 
\end{equation}
between the spherical harmonic coefficients of the surface load and those
of the resulting gravitational potential perturbation. Here, by use of the
appropriate loading Love number, we are accounting for the direct gravitational
effect of the surface mass along with the contribution to $\phi$ associated with
deformation of the earth model \citep[c.f.][eq.(12)]{wahr1998time}. We assume that the potential coefficients
at degrees zero and one vanish, the former due to conservation of mass and
latter because the measurements are made relative to
the centre of mass frame. For $l\ge 2$,
the Love numbers are non-zero and so we can write
\begin{equation}
  \sigma_{lm} = k_{l}^{-1}\,\phi_{lm}.
\end{equation}
and hence  we can  reconstruct the load through
\begin{equation}
  \label{eq:sigma_rec}
  \sigma \approx \sum_{lm} k_{l}^{-1}\,\phi_{lm}\,Y_{lm}, 
\end{equation}
where an approximately equals sign is included because terms for $l\le 1$ are excluded.
Errors in GRACE observations are known to increase quite rapidly as a function
of degree \citep[e.g.][Figure 1]{swenson2002methods}, and so
the summation in eq.(\ref{eq:sigma_rec}) must be truncated at some finite-degree, $L$,
with the the choice $L = 100$ being common. Using this approach one can obtain point-estimates of
surface loads from GRACE data. In fact, it is often more
useful to estimate average loads over a specific region of interest.
This is done by introducing an averaging function, $w$,
and then using eq.(\ref{eq:sigma_rec}) to obtain
\begin{equation}
  \label{eq:load_average}
  \int_{\partial M} w\,\sigma \dd S \approx \sum_{lm} b^{2} k_{l}^{-1}\,\phi_{lm} \,w_{lm}, 
\end{equation}
with $w_{lm}$ the expansion coefficients of the averaging function; in
what follows, we call the right hand side of this expression the ``GRACE estimate''
of the  load.  The  design of optimal averaging functions for a region has been discussed by
\cite{swenson2002methods}, but  in our numerical examples we only
use the simple Gaussian functions  of \cite{wahr1998time}.

From eq.(\ref{eq:load_average}), we can equivalently write the GRACE estimate as
\begin{equation}
\sum_{lm} b^{2} k_{l}^{-1}\,\phi_{lm} \,w_{lm}  =  \int_{\partial M} \tilde{w} \, \phi \dd S, 
\end{equation}
where we have defined
\begin{equation}
  \label{eq:zeta_phi_load_average}
  \tilde{w}  = \sum_{lm}  k_{l}^{-1}\, w_{lm}\, Y_{lm}, 
\end{equation}
with, again, the summation leaving implicit the inclusion of only degrees $ 2 \le l \le L$. In this manner, we can apply the reciprocity theorem stated in eq.(\ref{eq:rec7}) to write
\begin{equation}
  \label{eq:GRACE_kernel}
  \int_{\partial M} \tilde{w} \, \phi \dd S   = \int_{\partial M} \Delta SL^{\dagger} \, \zeta \dd S, 
\end{equation}
where $\Delta SL^{\dagger}$ is obtained by solving the generalised fingerprint problem subject to
the adjoint loads
\begin{equation}
  \zeta^{\dagger} = 0, \quad \zeta_{u}^{\dagger}(\mathbf{x})  = 0,
  \quad \zeta_{\phi}^{\dagger} = -g \,\tilde{w}, 
   \quad \mathbf{k}^{\dagger} = \int_{\partial M}\zeta_{\phi}^{\dagger} \, \mathbf{x} \times
    (\Bom \times \mathbf{x}) \dd S.
\end{equation}
The sensitivity kernel, $\Delta SL^{\dagger}$, in eq.(\ref{eq:GRACE_kernel})  quantifies the manner
in which  the GRACE estimate  depends on the direct load. Within regions
covered by grounded ice, the direct load $\zeta$ and the total load, $\sigma$,
are equal and hence  the difference between $SL^{\dagger}$ and $w$
determines the accuracy of the GRACE estimate. For example, if the
equality $SL^{\dagger} = w$ were to hold exactly  then 
\begin{equation}
  \int_{\partial M} \tilde{w} \, \phi \dd S = \int_{\partial M} w \, \zeta \dd S, 
\end{equation}
and hence, in the absence of observational errors, the GRACE estimate would
provide the exact average of the direct load  over the chosen region.
Differences between $\Delta SL^{\dagger}$ and $w$ are, therefore, indicative of systematic biases within
the GRACE estimate due to the neglect of indirect water loading.

To illustrate these ideas, we show in Fig.\ref{fig:GRACE_average_antarctica} (a) such an averaging
function targeted at West Antarctica. Specifically, we have used a Gaussian function
as defined as in eq.(32) of \cite{wahr1998time}, with its  centre at
($82^\circ$\,S, $110^{\circ}$\,W), and with  a half-width  of  $800$\,km. A value
of $-0.06378$\,m as an equivalent water thickness
was determined for the GRACE estimate  from our synthetic data set
assuming a truncation-degree of $L = 100$. This compares to the true value of
$-0.0692$\,m, meaning a relative error of $-7.86$\% associated with the method.
To understand this difference, we show in  Fig.\ref{fig:GRACE_average_antarctica} (b)  the sensitivity
kernel with respect to direct load for the estimate and in (c) the relative difference between
this kernel and the averaging function (with values normalised relative to the maximum value of the latter field).
While the relative difference between $w$ and $\Delta SL^{\dagger}$ is everywhere below about $2$\%, this
difference is non-zero over much of the Earth's surface, and it is this feature  that accounts for
the comparatively large error in the GRACE estimate. Analogous calculations
are shown in Fig.\ref{fig:GRACE_average_greenland} for an averaging function targeted
on the Greenland ice sheet; in detail the function is centred on  ($73^\circ$\,N, $40^{\circ}$\,W)
and has  a half-width of $700$\,km. In this case, the relative error in the GRACE estimate
was found to be $5.12$\% which is, again, significantly larger than the point-wise difference between
the averaging function and associated sensitivity kernel. The forward-modelling results
discussed here are consistent with those of  \cite{sterenborg2013bias},
but what is new  is the use of sensitivity kernels to  quantify more fully the
source of the bias. 

A point worth considering  is that, while GRACE  does not provide coefficients for $l\le1$,
the averaging function for the region of interest will
usually have some power at these degrees.  It has been shown \citep[e.g.][]{chambers2007effects}
that this can lead to significant biases within the load estimates, while efforts to constrain
the missing coefficients have been discussed using additional information
and assumptions \citep[][]{chambers2004preliminary,swenson2008estimating}.
It might, therefore, be reasonably asked if the errors seen  within our  GRACE estimates 
are  due predominantly to missing data at degrees $l \le 1$. If this were true, then the plots of
$\Delta SL^{\dagger}-w$ shown in sub-figure (c) of Fig.\ref{fig:GRACE_average_antarctica}
and Fig.\ref{fig:GRACE_average_greenland} should closely approximate $\bar{w}-w$ where
$\bar{w}$ denotes the projection of the averaging function onto degrees $l \ge 2$. The
latter fields are shown in sub-figure (d) of Fig.\ref{fig:GRACE_average_antarctica}
and Fig.\ref{fig:GRACE_average_greenland}, and here we  see clear differences
with those shown in (c), with the magnitude of the differences being comparable to that
of the underlying functions.

As a final remark on this example, within the traditional theory for estimating surface loads
from GRACE data \citep[e.g.][]{wahr1998time,swenson2002methods}, a direct
load at degree-one is associated with a degree-one gravitational potential perturbation, but such a perturbations
cannot be observed. A degree-one  direct load would, however, induce a water load that has power at higher degrees
and so \emph{is} associated with an observable gravitational signal. This suggests that, by more fully accounting
for the physics of the problem, it might
be possible to constrain degree-one loads
without recourse to additional observations or assumptions.

\subsection{Sensitivity kernels for sea surface height change}

\begin{figure}
  \centering
  \begin{subfigure}[c]{0.47\textwidth}
     \includegraphics[width=\textwidth]{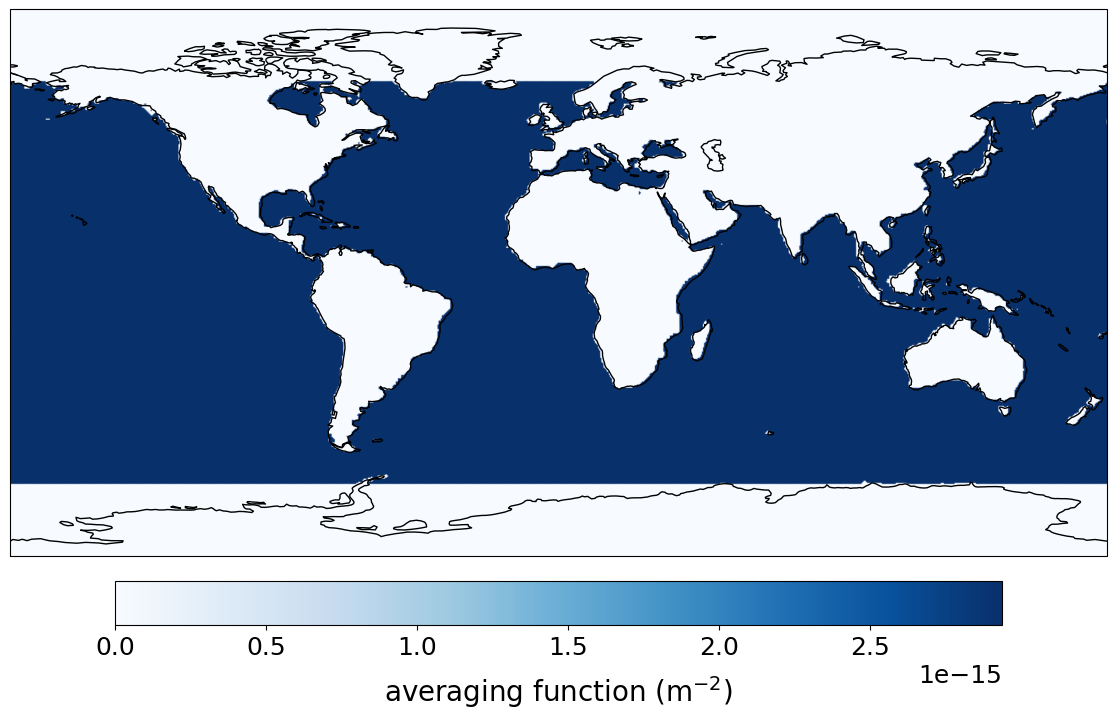}
     \caption{Altimetry averaging function}
   \end{subfigure} \hfill
   \begin{subfigure}[c]{0.47\textwidth}
     \includegraphics[width=\textwidth]{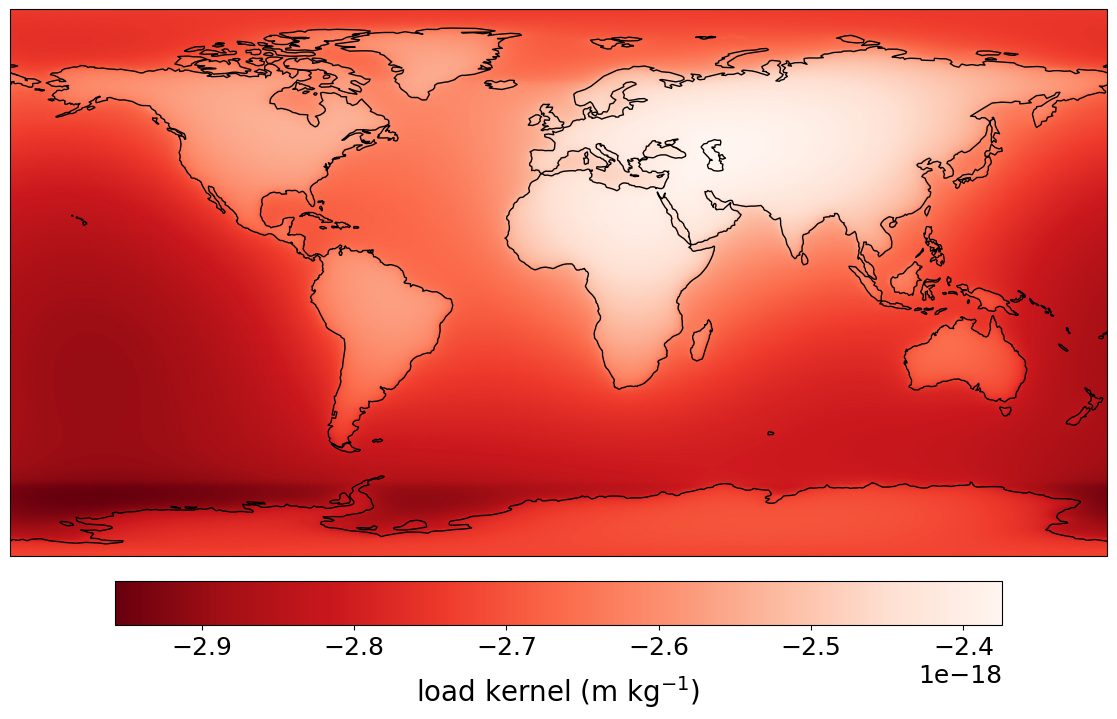}
     \caption{Load kernel}
   \end{subfigure}
  \begin{subfigure}[c]{0.47\textwidth}
     \includegraphics[width=\textwidth]{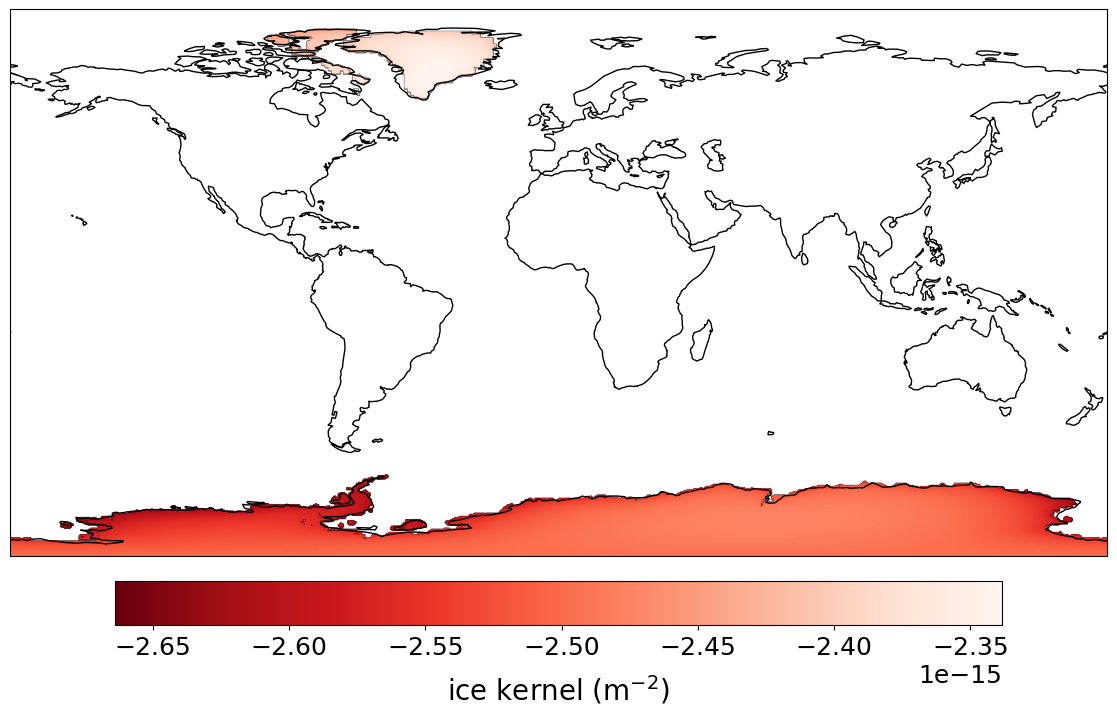}
     \caption{Ice kernel}
   \end{subfigure} \hfill
   \begin{subfigure}[c]{0.47\textwidth}
     \includegraphics[width=\textwidth]{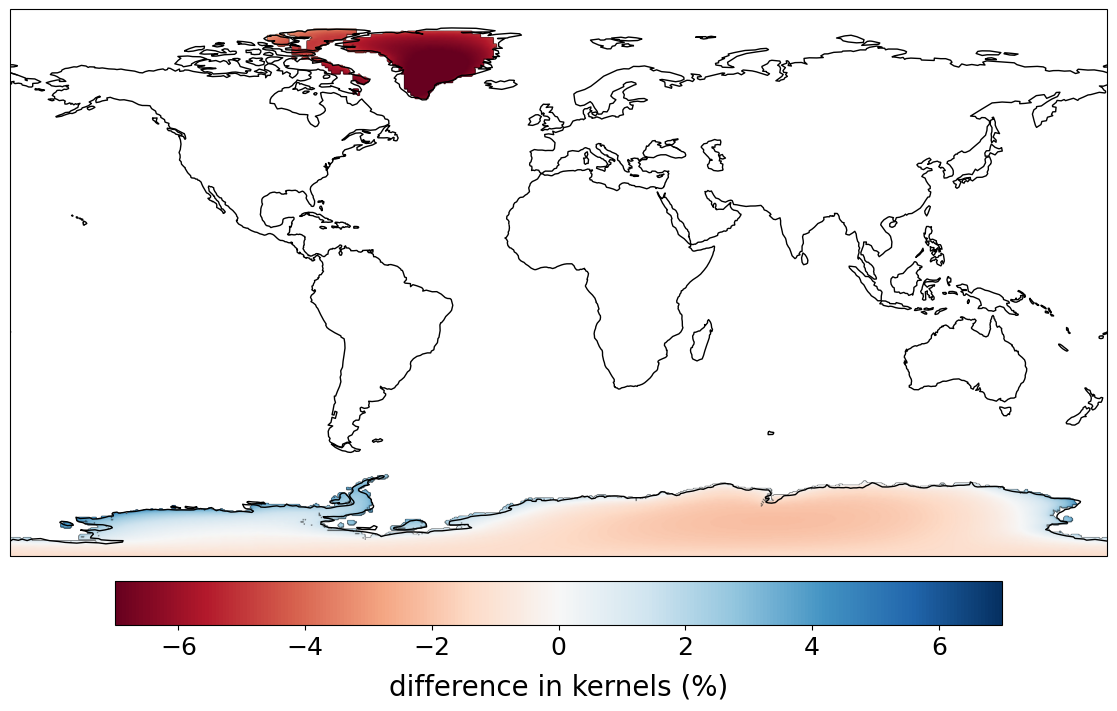}
     \caption{Difference with ideal kernel}
   \end{subfigure}
   \caption{
     In (a), we show the averaging function used to obtain an estimate of $\Delta GMSL$
     from sea surface height observations. This function is equal to a constant in the open oceans between
     $\pm 66^{\circ}$ in latitude, zero elsewhere, and integrates to one over the Earth's surface.
     The sensitivity kernel for such a measurement with respect to direct load is shown in (b)
     and with respect to ice thickness within glaciated regions in (c). The relative differences
     between the kernel in (c) and the  kernel that would return the exact value of
     $\Delta GMSL$ is shown in (d), with  values normalised relative
     to the latter kernel.  Regions in (d) where the difference is positive
     will have their contributions to the $\Delta GMSL$ estimate artificially amplified, and conversely where the
     difference is negative.}  
  \label{fig:altimetry}
\end{figure}

We now consider, in an idealised way, how global mean sea level change, $\Delta GMSL$, 
might be estimated from satellite altimetry data \citep[e.g.][]{nerem2010estimating,ablain2015improved}.
Suppose that we have measurements of sea surface height change, $\Delta SSH$, within the open oceans between
$\pm 66^{\circ}$ in latitude. An estimate of $\Delta GMSL$ can then be obtained by forming the average
\begin{equation}
  \Delta GMSL \approx \int_{\partial M} w\, \Delta SSH \dd S, 
\end{equation}
where $w$ is a function that has support  limited to regions where data is available;
we  will call the right hand side of this expression an ``altimetry estimate''.
Following \cite{ablain2015improved},  the averaging function we use is a constant within the
open oceans between latitudes $\pm 66^{\circ}$, zero elsewhere, and such that $\int_{\partial M } w \dd S = 1$; note that
their ``cosine of latitude weighting'' is provided by the surface element in the  integral.
To proceed, we recall \citep[e.g.][]{lickley2018bias} that the sea surface height change can be expressed
in the form
\begin{equation}
  \Delta SSH = \Delta SL + u + \frac{\psi}{g}.
\end{equation}
Note that here we are choosing to remove the centrifugal contribution to the sea surface
height change. Whether and how this should be done is a  subtle point related
to  reference frames that is discussed in detail by \cite{tamisiea2011ongoing} and \cite{lickley2018bias}.
For these illustrative purposes, however,
this choice does not matter appreciably. The altimetry averaging method
was applied to the synthetic data set shown in Fig.\ref{fig:forward_calculation}, leading to an
estimate of  $4.90$\,mm for global mean sea level change, this differing by $-1.99$\% from the true value of $5.00$\,mm.
By repeating such calculations for different melt geometries, it is possible to quantify   biases associated
with the method. For example, a second synthetic data set was generated in which melting
was limited to the northern hemisphere and, in this case, the relative error was larger at $-6.74$\%.
While such a forward modelling approach is undoubtedly useful, it does not provide a  clear means for
understanding the source of the systematic bias and hence is of limited utility in attempts to mitigate against  it.

Following the now familiar method, we can apply eq.(\ref{eq:rec6}) to obtain
\begin{equation}
  \int_{\partial M} w\, \Delta SSH \dd S = \int_{\partial M} \Delta SL^{\dagger} \,\zeta \dd S, 
\end{equation}
where $\Delta SL^{\dagger}$ is obtained through solution of the generalised fingerprint problem
subject to loads
\begin{equation}
\zeta^{\dagger} = w, \quad \zeta_{u}^{\dagger}(\mathbf{x})  = -w, 
\quad \zeta_{\phi}^{\dagger} = 0, \quad \mathbf{k}^{\dagger} = \int_{\partial M}\zeta^{\dagger} \, \mathbf{x} \times
    (\Bom \times \mathbf{x}) \dd S. 
\end{equation}
Within Fig.\ref{fig:altimetry}, we show the results of such a calculation for the
constant averaging function discussed above. Supposing that the sea level change is
driven entirely through mass transfer from the ice sheets, we can write
\begin{equation}
  \label{eq:GMSL}
  \Delta GMSL = -\frac{\rho_{i}}{\rho_{w} A}\int_{\partial M} (1-C)\,\Delta I \dd S.
\end{equation}
It follows that  the altimetry estimate would provide an exact
measure of $\Delta GMSL$ if its sensitivity kernel with respect to ice thickness
in glaciated regions was  equal the constant $-\frac{\rho_{i}}{\rho_{w} A}$.
From panel (c) within Fig.\ref{fig:altimetry}, we can see, however, that this kernel
undergoes spatial variations of order of $10$\%, with its value
in Greenland being uniformly lower than in Antarctica. To understand
the implications more clearly, we show in panel (d) the relative difference between
the ice kernel and the idealised kernel defined through
eq.(\ref{eq:GMSL}).  Within the northern hemisphere
the difference is largely uniform with value around $-6$\%.  In Antarctica, however, the difference between the
kernels is not of a uniform sign, with values of around $+2$\% found
in the west of the continent and mostly negative ones  around $-2$\% in the east.
These properties suggest that the altimetry averaging method  systematically underestimates northern
hemisphere contributions to global mean sea level, with this bias evident  within the
forward calculations discussed earlier within this subsection.

\section{Discussion}

Within this paper we have established a number of reciprocity theorems related to
surface loading and sea level change within an elastic earth model. This provides a simple method for
deriving and calculating sensitivity kernels for sea level and related observables with respect to changes in ice thickness.
While we have considered a range of different observables, these examples are not exhaustive. Using the
techniques developed it would be relatively easy  to obtain kernels for other relevant quantities
such as  absolute gravity \citep[e.g.][]{van2017using}, ice surface altimetry \citep[e.g.][]{remy2009antarctic} or horizontal component
GPS measurements \cite[e.g][]{coulson2021global}.
Our methods, moreover,  apply directly to other kinds of surface loading, such as that associated
with ocean dynamics, hydrological processes or sedimentation.

A range of example calculations have been presented to demonstrate the
validity of the theoretical results and to point to some potential applications. In future work, we will  show that these ideas,
when combined with modern geophysical inverse theory,  can be used to obtain better constraints on global mean
sea level  and ice sheet mass loss. To give an indication of how this might be done, we have seen  that estimates of global mean
sea level change obtained from uniform spatial averages of  satellite altimetry data suffer from systematic biases.
It is, however,  likely that some non-uniform average of the altimetry data will
do a better job. For any  proposed  average, we can use the tools developed in this paper to calculate the
associated sensitivity kernel, and hence quantify its performance as an estimator for global mean sea level change.
In this manner, the design of optimal estimators for specific quantities of interest can be undertaken
following well-established theoretical methods
\citep[e.g][]{backusgilbert,backus1970inferenceI,stark2008generalizing,stuart2010inverse}. 
Here, of course, it would be necessary
to account  for the full complexity of the real-world  problems including data errors and the need to
correct, albeit imperfectly, for physical process such as glacial isostatic adjustment and ocean dynamics
that affect the observations \citep[e.g.][]{chambers2010ocean,nerem2010estimating,hay2015probabilistic,kopp2015geographic,horton2018mapping}.

\begin{acknowledgments}
  DA and FS have been supported through the Natural Environment Research Council grant number NE/V010433/1.
  OC was supported by a NERC studentship and a CASE award from the British Antarctic Survey.
  JXM acknowledges support from  Harvard University and the MacArthur Foundation. AJL has been supported by the
  National Science Foundation under grants: NSF-EAR-2002352 and 1064 OPP-2142592.
\end{acknowledgments}

\section*{Data availability statement}

Codes needed to reproduce all calculations and figures  can be found at
https://github.com/da380/SLReciprocityGJI.git.

\bibliographystyle{gji}
\bibliography{references}

\end{document}